\documentclass[twocolumn,preprintnumbers,superscriptaddress,nofootinbib,aps,prd,floatfix]{revtex4}

\usepackage[toc,page,header]{appendix}

\usepackage{titletoc}

\usepackage{epsf,epsfig,subfigure,graphicx,amsmath,amssymb,cancel}
\usepackage{subfigure}
\usepackage{soul}
\usepackage[colorlinks=true,urlcolor=purple,anchorcolor=blue,citecolor=blue,filecolor=blue,linkcolor=red,menucolor=blue]{hyperref}
\usepackage[usenames,dvipsnames]{xcolor}
\usepackage{slashed}
\usepackage{bm}
\usepackage{enumitem}

\usepackage{afterpage}
\usepackage[utf8]{inputenc}
\usepackage{multirow}
\usepackage{makecell}

\allowdisplaybreaks

\newcommand{\newc}{\newcommand}
\newc{\gsim}{\lower.7ex\hbox{$\;\stackrel{\textstyle>}{\sim}\;$}}
\newc{\lsim}{\lower.7ex\hbox{$\;\stackrel{\textstyle<}{\sim}\;$}}
\newc{\gev}{\,{\rm GeV}}
\newc{\mev}{\,{\rm MeV}}
\newc{\ev}{\,{\rm eV}}
\newc{\kev}{\,{\rm keV}}
\newc{\tev}{\,{\rm TeV}}
\newc{\MHT}{$H_T^{\text{miss}}$}
\newc{\MET}{$\slashed{E}_T$}
\newc{\MTT}{$M_{T2}$}

\def\ln{\mathop{\rm ln}}

\def\Tr{\mathop{\rm Tr}}

\newc{\mz}{M_Z}
\newc{\mpl}{M_*}
\newc{\mw}{m_{\rm weak}}
\newc{\nr}[1]{N^c_R{}_{#1}}

\renewcommand{\a}{\alpha}

\newcommand{\bra}{\langle}
\newcommand{\ket}{\rangle}

\newcommand{\SU}{{\text{SU}}}
\newcommand{\up}{\uparrow}
\newcommand{\down}{\downarrow}

\def\beq{\begin{equation}}
\def\eeq{\end{equation}}
\newcommand{\bea}{\begin{eqnarray}\begin{aligned}}
\newcommand{\eea}{\end{aligned}\end{eqnarray}}
\def\bitem{\begin{itemize}}
\def\eitem{\end{itemize}}

\definecolor{darkgreen}{rgb}{0,0.6,0}
\definecolor{darkorange}{rgb}{0.99,0.5,0}

\usepackage{cleveref}
\crefname{table}{Table}{Tables}
\crefname{equation}{Eq.}{Eqs.}
\crefname{appendix}{App.}{Apps.}
\crefname{section}{Sec.}{Secs.}
\crefname{figure}{Fig.}{Figs.}

\usepackage{tikz}
\usetikzlibrary{decorations.pathreplacing}
\newcommand{\tikznode}[3][inner sep=0pt]{\tikz[remember
picture,baseline=(#2.base)]{\node(#2)[#1]{$#3$};}}
\usepackage{ytableau}

\usepackage[compat=1.0.0]{tikz-feynman}

\begin{document}


\vspace{-1cm}

\title{Noble Dark Matter: Surprising Elusiveness of Dark Baryons}

\vskip 1.0cm
\author{Pouya Asadi}
\thanks{{\scriptsize Email}: \href{mailto:pasadi@uoregon.edu}{pasadi@uoregon.edu}}
\author{Austin Batz}
\thanks{{\scriptsize Email}: \href{mailto:abatz@uoregon.edu}{abatz@uoregon.edu}}
\author{Graham D. Kribs}
\thanks{{\scriptsize Email}: \href{mailto:kribs@uoregon.edu}{kribs@uoregon.edu}}
\affiliation{Institute for Fundamental Science and Department of Physics, \\ University of Oregon, Eugene, OR, 97403, USA.}

\begin{abstract}

Dark matter could be a baryonic composite of strongly-coupled constituents 
transforming under SU(2)$_L$. We classify the SU(2)$_L$ representations of baryons in a class of simple confining dark sectors and find that the lightest state can be a pure singlet or a singlet that mixes with other neutral components of SU(2)$_L$ representations, which strongly suppresses the dark matter candidate's interactions with the Standard Model.
We focus on models with a confining $\SU(N_c)$ and heavy dark quarks constituting vector-like $N_f$-plet of $\SU(2)_L$. 
For benchmark $N_c$ and $N_f$, we calculate baryon mass spectra, incorporating electroweak gauge boson exchange in the non-relativistic quark model, and demonstrate that above TeV mass scales, dark matter is dominantly a singlet state. 
The combination of this singlet nature with the recently discovered $\mathcal{H}$-parity results in an inert state analogous to noble gases, 
hence we coin the term Noble Dark Matter. 
Our results can be understood in the non-relativistic effective theory that treats the dark baryons as elementary states, where we find singlets accompanying triplets, 5-plets, or more exotic representations.
This generalization of WIMP-like theories is 
more difficult to find or rule out than dark matter models that include only a single SU(2)$_L$ multiplet (such as a Wino), motivating new searches in colliders and a re-analysis of direct and indirect detection prospects in astrophysical observations.

\end{abstract}

\maketitle

\vskip 1cm

\renewcommand{\tocname}{\large Table of Contents}
 {
 \hypersetup{linkcolor=black}
 \tableofcontents
 }

\section{Introduction}
\label{sec:intro}

The existence of dark matter (DM) is well-established by gravitational observations across various distances, with no non-gravitational signals detected so far, see Ref.~\cite{Cirelli:2024ssz} for a recent review. 
On the one hand, such non-gravitational interactions of DM with the Standard Model (SM) are tightly constrained by direct and indirect detection searches. On the other hand,
nightmare scenarios of strictly gravitational interactions would not explain the similarity of DM and SM abundances.
These observations can be reconciled if a dark sector thermally interacted with the SM in the early universe before undergoing a phase transition that produced SM-singlet relics.

We show that this can occur in a broad class of confining dark sectors. 
We extend the SM by a sector of dark quarks charged under a confining $\SU(N_c)$ that form an $N_f$-plet of SM $\SU(2)_L$ with vanishing hypercharge, see \textit{e.g.}~Refs.~\cite{Kilic:2009mi,Bai:2010qg,Antipin:2014qva,Appelquist:2015yfa,Mitridate:2017oky,Kribs:2018oad,Kribs:2018ilo,Abe:2024mwa,Asadi:2024bbq} for previous studies of similar theories. 
The lightest dark baryons in the confined phase are stable and, if neutral, constitute a viable DM candidate.
This is a simple DM scenario where the dark sector was in thermal contact with the SM in the early universe (without invoking an additional portal), while the degrees of freedom present today can have very feeble SM interactions, depending on the values of $N_c$ and $N_f$.

In the deconfined phase of the theory, this model contains ingredients 
similar to the standard Weakly Interacting Massive Particle (WIMP), with the addition of a dark confining gauge group. 
In the confined phase of the theory, the dark quarks assemble into dark hadrons, including a WIMP-like dark baryon DM candidate. 
Similar to a vanilla WIMP model, the dark baryons transform in particular representations of SU(2)$_L$.
Unlike a vanilla WIMP model, we remain agnostic to the mechanism that fixes the DM relic abundance in the early universe, focusing on the prospects of different discovery channels today.

Regardless of $N_{c}$ and $N_{f}$, the DM candidate has suppressed electromagnetic interactions with SM due to $\mathcal{H}$-parity \cite{Asadi:2024bbq}.
This symmetry forbids the leading electromagnetic moments of the dark sector's neutral hadrons, which strongly suppresses direct detection signals in our model. 
Therefore, the leading direct detection signals arise from electroweak loop-induced interactions due to the  renormalizable interactions of the dark baryons with the $W$ gauge boson \cite{Essig:2007az,Hisano:2010fy,Hill:2014yka,Hill:2014yxa,Chen:2018uqz,Bottaro:2021snn,Bottaro:2022one,Chen:2023bwg,Bloch:2024suj}. For baryons in the singlet representation of $\SU(2)_L$, these interactions vanish as well.

We study the $\SU(2)_L$ representations of dark baryons for various combinations of $N_c$ and $N_f$ using standard group theory arguments. 
In the specific case where $N_f=2$ and $N_c$ is even, the lightest dark baryon is a pure singlet,\footnote{By ``singlet", we refer specifically to a state transforming in the trivial representation of SU(2)$_L$. Neutral components of non-trivial SU(2)$_L$ multiplets are referred to as such.} and the interactions with the SM become highly suppressed.\footnote{The leading interaction is the electromagnetic polarizability of the composite singlet.} 
For many other combinations of $N_c$ and $N_f$, we find that the DM state is composed primarily of a singlet of SU(2)$_L$ with only a small mixing with non-singlet dark baryons, so the elastic scattering rate relevant to direct detection signals of these models is suppressed by this small mixing.
These (approximate) singlet baryons are inert composites of charged constituents, analogous to noble gases, so we have named them Noble Dark Matter.

Despite confining dark sectors being targets of different search strategies, \textit{e.g.} see Refs.~\cite{Strassler:2006im,Han:2007ae,Kang:2008ea,Juknevich:2009ji,Kilic:2009mi,Kribs:2009fy,Juknevich:2009gg,Harnik:2011mv,Frigerio:2012uc,Schwaller:2015gea,Cohen:2015toa,Appelquist:2015yfa,Carmona:2015haa,Knapen:2016hky,Knapen:2017kly,Kribs:2018ilo,Kribs:2018oad,Evans:2018jmd,Bernreuther:2019pfb,Bernreuther:2020vhm,Knapen:2021eip,Barron:2021btf,Kuwahara:2023vfc,Batz:2023zef,Bagnasco:1993st,Alves:2009nf,SpierMoreiraAlves:2010err,Buckley:2012ky,Bhattacharya:2013kma,Antipin:2015xia,Hardy:2015boa,DeLuca:2018mzn,Contino:2020god,Bloch:2024suj,Detmold:2014qqa,Soni:2016gzf,Mahbubani:2019pij,Cline:2013zca,Boddy:2014yra,Krnjaic:2014xza,Buen-Abad:2015ova,Gross:2018zha,Morrison:2020yeg,Contino:2020tix,Garani:2021zrr}, our results imply that even this simplest confining extension of the WIMP paradigm is far more elusive than historically appreciated (see also Ref.~\cite{Krall:2017xij} for non-confining WIMPs that are notoriously difficult to detect). Confinement in the dark sector naturally gives rise to a rich spectrum of WIMP-like states in the infrared (IR) despite including only one set of dark quarks in the ultraviolet (UV) that transform in a vector-like representation of the electroweak and the dark gauge groups.
The effective field theory (EFT) in the IR inherits symmetries from the UV, particularly $\mathcal{H}$-parity, that forbid some interactions between the dark sector and the SM. The UV theory also predicts other mechanisms that suppress interactions between the SM and the dark sector in the IR (\emph{i.e.}~singlets with small mixings with other states) that may appear unnatural or contrived from a bottom-up perspective. Earlier simplified model approaches introducing a single (possibly composite) electroweak multiplet have neglected this possibility.

The structure of the paper is as follows:  In \cref{sec:model}, we introduce our model, review $\mathcal{H}$-parity, and explain baryon SU(2)$_L$ representations. 
We compute the lowest-lying baryon mass spectra and mixings of states in different $\SU(2)_L$ representations for two $N_{c,f}$ combinations in \cref{sec:spectrum} using the non-relativistic quark model, including electroweak contributions. 
We find that the DM candidate in these examples is mostly made of an $\SU(2)_L$ singlet state and discuss in \cref{sec:pheno} the resulting suppression of (in)direct detection signals, motivating searches at current and future high-energy colliders. We conclude in \cref{sec:conclusion}.
In \cref{app:running}, we study the effect of the new fields on the running of SU(2)$_L$ gauge coupling. Further details on group theory arguments relevant for determining the SU(2)$_L$ representations and wavefunctions of dark baryons are provided in \cref{app:group_theory}. 
We also discuss details of a variation method used in determining the baryon mass spectrum in \cref{app:mass}.

\section{The Theory}
\label{sec:model}

\subsection{\texorpdfstring{$\mathcal{H}$}{H}-parity from the Ultraviolet}
\label{subsec:UV}

We propose a theory of dark matter that arises from a new strongly-coupled confining gauge sector
with the Lagrangian

\begin{equation}
\label{eq:L}
  \mathcal{L}_{\text{dark}} = -\frac{1}{4} G^{\mu\nu\,a}G_{\mu\nu}^a
  + \overline{\mathbf{Q}} \left(i\slashed{D}
  - m_0 \right)\!\mathbf{Q}\,,
\end{equation}

\noindent where $G_{\mu\nu}^{a}$ is the field strength of the dark $\SU(N_c)$ with confinement scale $\Lambda_\chi$ and coupling $\a_\chi$, and $\mathbf{Q}$ is a vector-like dark quark with bare mass $m_0$ that transforms as a fundamental of $\SU(N_c)$ and an $N_f$-plet of $\SU(2)_L$.\footnote{The added matter content can spoil the asymptotic freedom of $\SU(2)_L$. See \cref{app:running} for a discussion of Landau poles.} 
In the confined phase of the theory, dark quarks bind into  mesons and baryons. 
Since $\mathcal{L}_{\text{dark}}$ includes no hypercharge interactions, each quark or hadron in the dark sector has an electric charge equal to its weak isospin. 

The lightest baryon in the confined phase is stabilized by a U$(1)_{\text{baryon}}$ in the standard way. 
If this baryon is neutral, it can be a viable DM candidate.
As shown in Ref.~\cite{Asadi:2024bbq}, electromagnetic 
moments of the neutral dark baryons are strongly suppressed because they are eigenstates of the $\mathcal{H}$-parity transformation (where the electroweak gauge fields are charge-conjugated and $\mathbf{Q}\to e^{i\pi J_y}\mathbf{Q}$, with $J_y$ the second generator of $\SU(2)_L$). 
In the upcoming sections, we discuss the baryons' SU(2)$_L$ representations and mass spectrum, as well as the phenomenological implications of our findings. The vanishing of baryon electromagnetic moments and the spectrum of SU(2)$_L$ representations have a strong impact on possible experimental signatures.

\subsection{Baryon Representations in the Infrared}
\label{subsec:IR}

In the confined phase of the theory, hadron states can be organized into multiplets transforming in irreducible representations of SU(2)$_L$.
Baryon wavefunctions can be thought of as multi-particle states of identical fermions tracking each particle's position, dark color, flavor, and spin. For symmetric spatial states and anti-symmetric (SU($N_c$)-singlet) color states, the required constituent exchange symmetry of baryon spin-flavor wavefunctions restricts baryon states to those in specific flavor representations of $\SU(N_f)$. In particular, the allowed flavor representations of $\SU(N_f)$ have a corresponding $\SU(2)_{\text{spin}}$ representation with an identical Young tableau. These tableaux have the form

\begin{gather} \label{eq:NfrepNoD}
\begin{ytableau}
       \tikznode{a1}{~} & \none[\cdots] & \tikznode{a2}{~} & \tikznode{a3}{~} & \none[\cdots] & \tikznode{a4}{~} \\
       \tikznode{b1}{~} & \none[\cdots] & \tikznode{b2}{~} & \none & \none & \none
\end{ytableau} \\ \notag
\end{gather}
\tikz[overlay,remember picture]{
\draw[decorate,decoration={brace}] ([yshift=-3mm,xshift=2mm]b2.north east) -- 
([yshift=-3mm,xshift=-2mm]b1.north west) node[midway,below]{$N_c/2-S$};
}
\tikz[overlay,remember picture]{
\draw[decorate,decoration={brace}] ([yshift=-3mm,xshift=2mm]a4.north east) -- 
([yshift=-3mm,xshift=-2mm]a3.north west) node[midway,below]{$2S$};
}

\noindent for spin-$S$ baryons because $N_c$ dark quarks are combined to make a color-singlet baryon, and the tableau for a spin-$S$ representation of $\SU(2)_{\text{spin}}$ has at most two boxes in each column and $2S$ columns with one box. See \cref{subsec:spin_flav_reps} for more details.

Enumerating the dark baryons' $\SU(2)_L$ representations amounts to decomposing the $\SU(N_f)$ representation in \cref{eq:NfrepNoD} into representations of its gauged $\SU(2)$ subgroup. We explain the derivation of these decompositions and hadron spin-flavor wavefunctions in \cref{subsec:algorithm}. 

In the electroweak-broken phase, states in different $\SU(2)_L$ representations can mix. However, neutral baryons in adjacent odd representations have opposite charges under $\mathcal{H}$-parity \cite{Asadi:2024bbq}, which forbids their mixing. For example, a singlet can mix with the neutral components of 5-plets, 9-plets, \emph{etc.}~($\mathcal{H}$-parity even), while it cannot mix with neutral components of 3-plets, 7-plets, \emph{etc.}~($\mathcal{H}$-parity odd).

\begin{table}[t]
    \centering
    \resizebox{\columnwidth}{!}{
    \begin{tabular}{c|c|c} \hline
        ($N_c$,~$N_f$) & Baryon SU(2)$_L$ Representations & Category \\
        \hline
         (2,2) & $\bm{1}$ & I \\
         \hline
         (2,3) & $\bm{3}$ & III \\
         \hline
         (2,4) & $\bm{5} \oplus \bm{1}$ & II \\
         \hline
         (2,5) & $\bm{7} \oplus \bm{3} $ & III \\
         \hline
         \hline
         (3,2) & $\bm{2}$ & IV \\
         \hline
         (3,3) & $\bm{5} \oplus \bm{3} $ & III \\
         \hline
         (3,4) & $\bm{8} \oplus \bm{6} \oplus \bm{4} \oplus \bm{2}$ & IV \\
         \hline
         (3,5) & $\bm{11} \oplus \bm{9} \oplus \bm{7} \oplus \bm{5} \oplus \bm{5} \oplus \bm{3} $ & III \\
         \hline
         \hline
         (4,2) & $\bm{1}$ & I \\
         \hline
         (4,3) & $\bm{5} \oplus \bm{1}$ & II \\
         \hline
         (4,4) & $\bm{9} \oplus \bm{5} \oplus \bm{5} \oplus \bm{1}$ & II \\
         \hline
         \multirowcell{1}{(4,5)} & \multirowcell{1}{$\bm{13} \oplus \bm{9} \oplus \bm{9} \oplus \bm{7} \oplus \bm{5} \oplus \bm{5} \oplus \bm{1} \oplus\bm{1}$} & \multirowcell{1}{II} \\
         \hline
         \hline
         (5,2) & $\bm{2}$ & IV \\
         \hline
         (5,3) & $\bm{7} \oplus \bm{5} \oplus \bm{3} $ & III \\
         \hline
         (5,4) & $\bm{12} \oplus \bm{10} \oplus \bm{8} \oplus \bm{8} \oplus \bm{6} \oplus \bm{6} \oplus \bm{4} \oplus \bm{4} \oplus \bm{2}$ & IV \\
         \hline
         \multirowcell{3}{(5,5)} & \multirowcell{3}{$\bm{17} \oplus \bm{15} \oplus \bm{13} \oplus \bm{13} \oplus \bm{11} \oplus \bm{11} $ \\
         $ \oplus\, \bm{11} \oplus \bm{9} \oplus \bm{9} \oplus \bm{9} \oplus \bm{9} \oplus \bm{7} \oplus \bm{7} \oplus \bm{7}$ \\
         $ \oplus \,\bm{5} \oplus \bm{5} \oplus \bm{5} \oplus \bm{5} \oplus \bm{3} \oplus \bm{3} \oplus \bm{1}$}
         & \multirowcell{3}{II} \\
         & & \\
         & & \\ \hline
    \end{tabular}
    }
    \caption{$\SU(2)_L$ representations of the lowest-spin dark baryons in the confined phase of the model in \cref{eq:L} for several combinations of number of dark colors $N_c$ and flavors $N_f$. 
    We define in the text the four categories in the rightmost column describing when singlets and other neutral states appear.}
    \label{tab:money_table}
\end{table}

In \cref{tab:money_table}, we show the $\SU(2)_L$ representations of all lowest-spin and spatially symmetric baryons for each set of $N_{c,f}=2,\ldots,5$ (see \cref{subsec:more_reps} for more). 
Note the appearance of SU(2)$_L$ singlets, as states composed of singlets will have 
important implications for dark matter interactions, as we will see.
We can classify the theory space into four categories:\footnote{Analogous categories exist when the quarks are spin-0, and singlets appear with similar frequency. See \cref{subsec:squarks}.}
\begin{enumerate}[label=\Roman*.]

    \item When $N_c$ is even and $N_f=2$, the lowest-spin baryon spectrum includes only a singlet of SU(2)$_L$, as one can infer from \cref{eq:NfrepNoD}.

     \item One or more singlets are accompanied by non-trivial odd representations of SU(2)$_L$. The lightest state in the spectrum can be a mixture of the singlet and other neutral states.

    \item There are no singlets, and instead only non-trivial odd representations of SU(2)$_L$.
    The lightest baryon may be neutral, but it must transform non-trivially under SU(2)$_L$.
    
    \item When $N_c$ is odd and $N_f$ is even, the baryons are in even representations of SU(2)$_L$, so none of them are neutral.
    
\end{enumerate}

\noindent We discuss the phenomenology of prospective DM candidates corresponding to each category in \cref{sec:pheno}. As we will explain, categories I and II are particularly interesting to us.

\section{Masses in the Heavy Quark Limit}
\label{sec:spectrum}

Our DM candidate is the lightest mass eigenstate in the dark baryon spectrum. To identify which baryon state is the lightest, we use the non-relativistic quark model in the heavy quark limit, where hadrons are well-approximated as weakly-coupled bound states of their valence quarks \cite{DeRujula:1975qlm, Manohar:1983md, Georgi:1984zwz}. We discuss the mass Hamiltonian in this limit and our approximation of its ground state using the variational method. We focus on the models with $(N_c,N_f) = (2,4)$ and $(4,3)$, as these are the simplest cases in category II defined in \cref{subsec:IR} (where the baryon spectra include singlets among neutral states in non-trivial SU(2)$_L$ multiplets). 

\subsection{Electroweak Contributions in the Quark Model}
\label{subsec:EWinQM}

Including leading-order $\SU(N_c)$ and electroweak effects, the spatially-symmetric baryon mass is given by the operator (derived in \cref{app:Hamiltonian}) \cite{DeSanctis:2009zz}

\begin{align} \label{eq:M}
M &= \sum_{i}\left(
m_i + \frac{\left\bra p_i^2\right\ket}{2m_i}
\right) \\
& \ \ \ + \sum_{i<j} \left\bra
V^\chi_{ij} + V^{\text{EM}}_{ij} + V^Z_{ij} + V^W_{ij}
\right\ket + \mathcal{O}(p_{i,j}^3/m_q^3), \notag \nonumber \\
\bra V^\chi_{ij} \ket &= -\alpha_\chi\frac{N_c+1}{2N_c}\  \mathcal{V}(0) + \mathcal{O}(\alpha_\chi^2),  \nonumber \\
\bra V^{\text{EM}}_{ij} \ket &= \alpha_{\text{EM}}Q_iQ_j \mathcal{V}(0) 
+ \mathcal{O}(\alpha_{\text{EM}}^2), \nonumber \\
\bra V^{Z}_{ij} \ket &= \alpha_{W}\cos^2\theta_WQ_iQ_j \mathcal{V}(m_Z) + \mathcal{O}(\alpha_W^2),  \nonumber \\
\bra V^{W}_{ij} \ket &= \frac{\alpha_{W}}{2}  \left(
J_{+}^iJ_{-}^j+J_{-}^iJ_{+}^j
\right)\mathcal{V}(m_W)
+ \mathcal{O}(\alpha_W^2),  \nonumber \\
\mathcal{V}(m_V) &= a_1 + \frac{1}{4m_q^2}\left(
1+\frac{8}{3} {\bm S}_i\cdot{\bm S}_j
\right) (m_V^2 a_1-4\pi \delta) \notag \nonumber \\
&\ \ \ + \frac{m_V}{8m_q^2}\left(
m_V^2a_0-6m_Va_1+4a_2
\right) \nonumber \\
&\ \ \ -\frac{1}{4m_q^2}\left(
m_V^2c_1 -4m_Vc_2-2b+m_Vd_2-2d_3
\right),\notag \nonumber 
\end{align}

\noindent where sums are over constituents in the $i^{\text{th}}$ position of the spin-flavor wavefunction with masses $m_i$, momenta ${\bm p}_i$, spins ${\bm S}_i$, and charges $Q_i$. The inter-quark potential has terms from exchange of a dark gluon ($V^\chi_{ij}$), photon ($V^{\text{EM}}_{ij}$), and weak bosons ($V^{W,Z}_{ij}$). 
The electromagnetic and $\SU(2)_L$ couplings $\alpha_{\text{EM}}$ and $\alpha_{W}$ appear, as well as the Weinberg angle $\theta_W$. 
$J^i_\pm$ are $\SU(2)_L$ ladder operators acting on the $i^{\text{th}}$ constituent (see \cref{subsec:algorithm} for explicit forms). Expectation values $\langle \cdot \rangle$ in \cref{eq:M} are taken with respect to spatial wavefunctions, so $M$ is an operator that acts on baryon spin-flavor states. The mass $m_q$ is the common tree-level mass of the quarks, which is distinct from $m_i$ due to self-energy corrections inducing quark mass splittings (see \cref{app:quark_splittings}). The function $\mathcal{V}$ depends on the vector boson mass $m_V$ and various spatial expectation values, written in the position basis as

\begin{align}
\label{eq:fspatial}
a_n &= \left\bra r^{-n}e^{-m_Vr} \right\ket , \\
b &= \left\bra r^{-1}e^{-m_Vr}\,{\bm \nabla}_i \cdot {\bm \nabla}_j \right\ket , \nonumber \\
c_n &= \left\bra r^{-n}e^{-m_Vr} {\bm r}\cdot ({\bm \nabla}_i - {\bm \nabla}_j) \right\ket ,\nonumber \\ 
d_n &= \left\bra r^{-n}e^{-m_Vr}{\bm r}\cdot ({\bm r}\cdot{\bm \nabla}_i)  {\bm \nabla}_j \right\ket, \nonumber \\
\delta &= \left\bra \delta^{(3)}({\bm r}) \right\ket ,\nonumber 
\end{align}

\noindent with ${\bm r} = {\bm r}_i - {\bm r}_j$ the inter-quark separation and ${\bm \nabla}_i$ the gradient with respect to ${\bm r}_i$. The potential is a generalization of the well-known Fermi-Breit potential (used in Ref.~\cite{DeRujula:1975qlm} for the electromagnetic and strong potentials in the SM) to include effects of the $W$ and $Z$ masses $m_{W,Z}$.

The $W$ and $Z$ potentials include contact terms (with the $\delta$ factor) that are unsuppressed in the limit $m_q/m_V \rightarrow 0 $. These terms are essential for $M$ to be consistent with gauge invariance in the limit $m_{W,Z}\to0$, as explained in \cref{subsec:sfmatrix}. Also, new terms arise in the finite-$m_{W,Z}$ regime that are not present in the usual Fermi-Breit potential for massless vector boson exchange \cite{DeSanctis:2009zz}. 
We must account for both of these effects to accurately capture the mass splittings induced by differences among baryon spin-flavor matrix elements of operators with $Q_iQ_j$ and $J^i_\pm J^j_\mp$. 

There are two physically-relevant bases in which we express baryon states: the SU(2)$_L$ basis and the mass basis. In the SU(2)$_L$ basis, each state is a component of a multiplet that transforms in a well-defined irreducible representation of SU(2)$_L$. We express the mass operator in \cref{eq:M} in the SU(2)$_L$ basis, then diagonalize it to transform to the mass basis. In this way, we can express a baryon mass eigenstate as a linear combination of states with well-defined SM interactions.

\subsection{Mass Spectra from the Variational Method}
\label{subsec:result_spectrum}

To estimate the spatial expectation values in \cref{eq:fspatial}, we use the variational method with the following template spatial wavefunction:

\begin{equation}\label{eq:trial_psi}
\psi = \left(1 + k_1 \sum_{i<j} |\bm{r}_i - \bm{r}_j| \right)e^{-k \sum_{i<j}|\bm{r}_i - \bm{r}_j|}\,,
\end{equation}

\noindent where $k$ and $k_1$ are optimization parameters with respect to which we minimize mass eigenvalues. The function is symmetric in the exchange of constituents, as we are interested in the lowest-lying ($s$-wave) hadrons. Similar template functions were used in Ref.~\cite{Mitridate:2017oky}, and an overall exponential suppression is typical in other weakly-bound states such as atoms. Further details on how we compute the spatial expectation values are in \cref{subsec:spatial}.

\begin{figure} [t]
            \resizebox{0.99\columnwidth}{!}{
            \includegraphics[width=\linewidth]{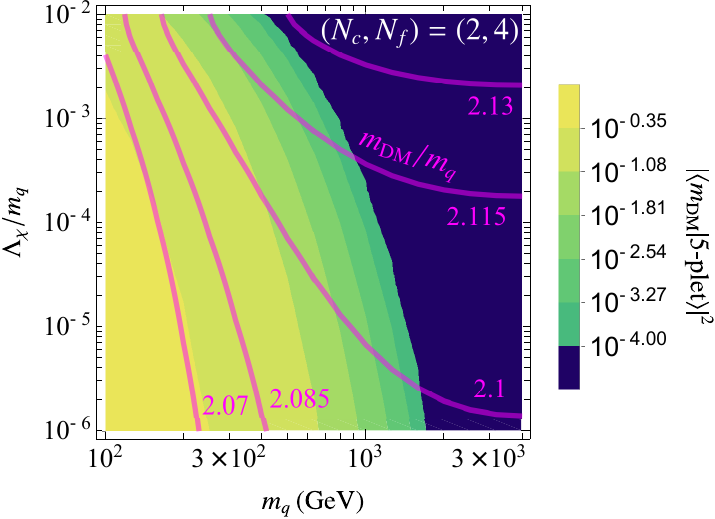}
            }
            \resizebox{\columnwidth}{!}{
            \includegraphics[width=\linewidth]{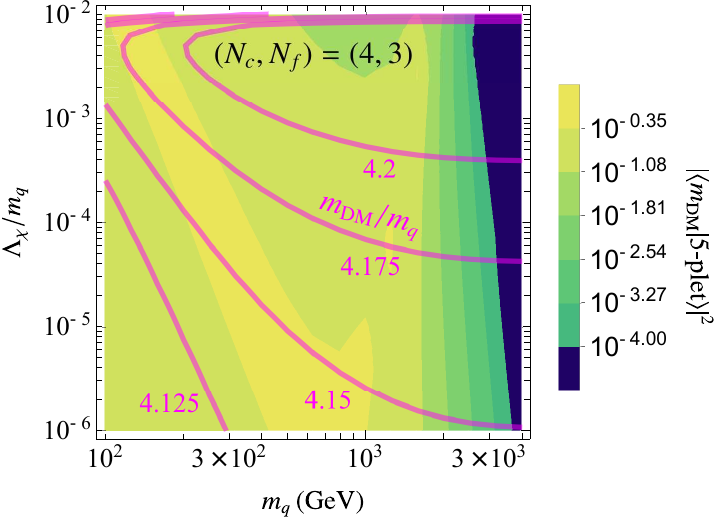}
            }
            \caption{Mass $m_{\text{DM}}$ of the lightest baryon and overlap of the lightest mass eigenstate $|m_{\text{DM}}\ket$ with the neutral $\SU(2)_L$ 5-plet state $|\text{5-plet}\ket$ for $(N_c,N_f)=(2,4)$ (\textbf{top}) and $(4,3)$ (\textbf{bottom}), which are two benchmark models from category II in \cref{subsec:IR}.
            As the dark quark mass $m_q$ increases, DM's overlap with the non-singlet state decreases. 
            DM masses are slightly greater than $N_c m_q$ due to quark self-energy corrections.  }
            \label{fig:mdm}
\end{figure}

We show the masses of the lightest neutral baryons and their overlaps with the $\SU(2)_L$ 5-plet states in \cref{fig:mdm} for $(N_c,N_f)=(2,4)$ and $(4,3)$. 
In the full parameter space shown, the lightest baryons are neutral and therefore may be viable DM candidates. 
The numerous terms in \cref{eq:M}, each with different dependence on $m_q$ and $\Lambda_\chi$, give rise to non-trivial features in the mixing contours.

We verified that our calculation satisfies various consistency checks. For example, the quark-level calculation should reproduce expectations from hadron-level self-energy corrections. In particular, charged hadrons should typically be heavier than neutral hadrons, and hadron states in $\SU(2)_L$ representations with larger Casimir factors should have a larger overlap with heavier mass eigenstates. Indeed, we find that the lightest baryon is neutral in the full parameter space shown, and the 5-plet has larger overlap with heavier mass eigenstates than the singlet. The heavier neutral mass eigenstate can be heavier than the singly-charged state in the small-$m_q$ region due to quark mass splittings or differences in electroweak matrix elements. See \cref{subsec:sfmatrix} for elaboration.

In our numerical evaluation of the mass spectrum, we verified that in the limit of $m_q,\Lambda_\chi \gg m_{W,Z}$, the mixing between the singlet and the neutral component of the 5-plet vanish, as expected.
In the large-$m_q$ regime, we find the mass splittings between the heavier neutral and singly-charged baryons to be $\mathcal{O}(10^2)\,\text{MeV}$, which is very close to the $166\;\text{MeV}$ expectation for the mass splitting of any heavy singly-charged and neutral component of an $\SU(2)_L$ multiplet \cite{Cirelli:2005uq}.

\section{Phenomenology}
\label{sec:pheno}

In this section, we discuss the phenomenology of our Noble Dark Matter in different contexts. In particular, we highlight their elusiveness in direct detection experiments, while also commenting on possible collider and astrophysical signatures.

\subsection{Direct Detection}
\label{subsec:DD}

As first described in Ref.~\cite{Asadi:2024bbq}, $\mathcal{H}$-parity forbids leading electromagnetic moments of neutral hadrons in our model for all $N_{c,f}$ values, and thus significantly reduces the cross section of $\mathcal{H}$-parity-symmetric DM with the SM.
Therefore, as with WIMPs, electroweak loop-induced interactions become the dominant source of direct detection signal \cite{Essig:2007az,Hisano:2010fy,Hill:2014yka,Hill:2014yxa,Chen:2018uqz,Bottaro:2021snn,Bottaro:2022one,Chen:2023bwg,Bloch:2024suj}. 
The strength of these interactions strongly depends on the lightest baryon's representation under SU(2)$_L$, giving rise to different phenomenology in each of the four categories defined in \cref{subsec:IR}. 
Below, we discuss the direct detection signal of each category.

\begin{enumerate}[label=\Roman*.]

    \item DM is a pure SU(2)$_L$ singlet,\footnote{For the specific case of $N_{c,f}=2$, it is shown that bound states of dark baryons can be stable and constitute a fraction of DM \cite{Detmold:2014qqa,Detmold:2014kba}.} so it does not have renormalizable interactions with electroweak gauge bosons. 
    This is nearly a direct detection nightmare scenario.\footnote{The leading direct detection signal becomes polarizability, to which foreseeable experiments are insensitive for DM masses $>\mathcal{O}(100)\,\text{GeV}$ \cite{Asadi:2024bbq}.}

    \item DM can be a mixture of the singlet and other neutral states, as shown in \cref{fig:mdm} for $(N_c,N_f)=(2,4)$ and $(4,3)$.
    Electroweak loop-induced signals are suppressed by this mixing, as explored below. 

    \item DM can be a mixture of neutral states in different non-trivial SU(2)$_L$ multiplets, and their interfering electroweak loop-induced interactions would need to be calculated carefully. In the special case where there is only one multiplet, or if the mixing of the lightest state with neutral states in other multiplets is forbidden by $\mathcal{H}$-parity,\footnote{An example of this case is $(N_c,N_f)=(3,3)$ (studied in Ref.~\cite{Asadi:2024bbq}). The lowest-spin baryon spectrum contains only a 3-plet and 5-plet (adjacent odd representations), so $\mathcal{H}$-parity forbids the mixing of the neutral components.} then DM has well-defined non-trivial interactions with the $W$ (similar to the standard WIMP).
    
    \item  When there are no neutral baryons, there is no viable single-baryon DM candidate. We do not discuss this category further in this work. 
    
\end{enumerate}

To constrain categories II and III with direct detection experiments, one should first identify the lightest mass eigenstate, verify it is neutral, and compute its composition of states in each $\SU(2)_L$ representation. In the heavy quark limit, this can be carried out with the variational method we proposed in \cref{sec:spectrum}.

\noindent 

\begin{figure} [t]
            \resizebox{\columnwidth}{!}{
            \includegraphics[width=\linewidth]{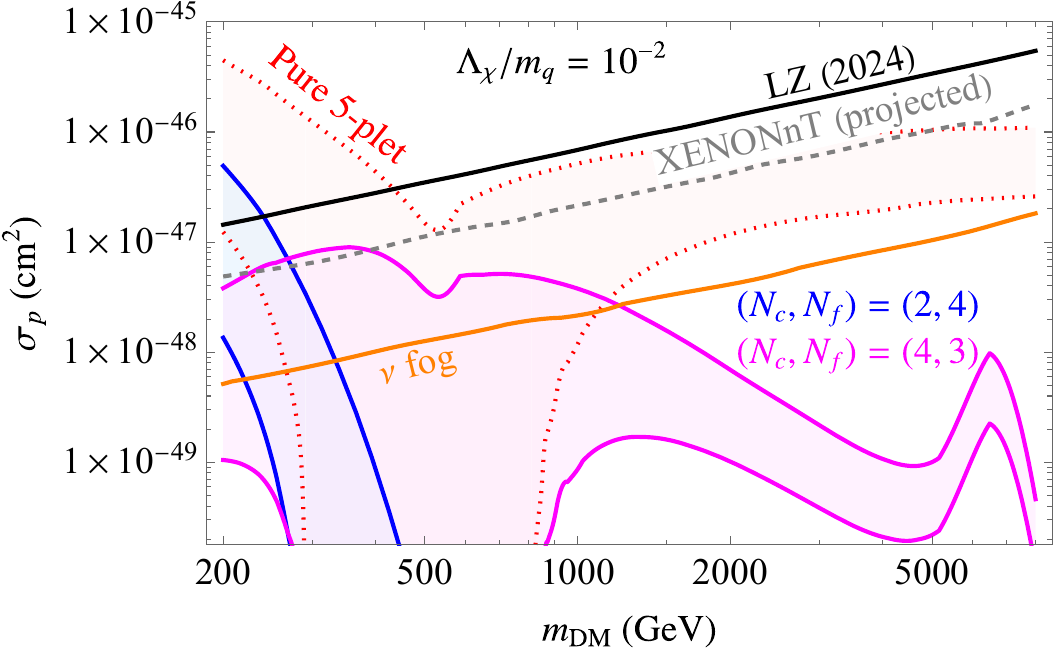}
            }\\
            \resizebox{\columnwidth}{!}{
            \includegraphics[width=\linewidth]{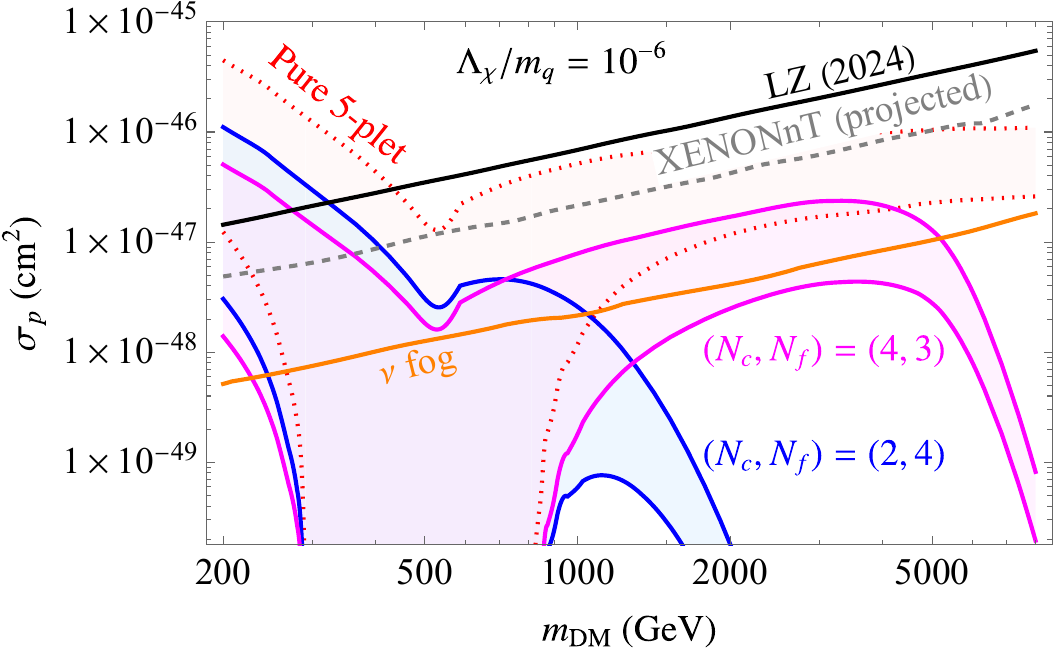}
            }
            \caption{Rough estimates of direct detection cross sections of the lightest dark baryon for $(N_c,N_f)=(2,4)$ and $(4,3)$ with  $\Lambda_\chi/m_{q}=10^{-2}$ (\textbf{top}) and $10^{-6}$ (\textbf{bottom}). We scaled the $\SU(2)_L$ 5-plet cross sections (red dotted line) computed in Ref.~\cite{Chen:2023bwg} using a similar 5-plet WIMP model by $\left|\left\langle m_{\text{DM}} | 
            5\text{-plet}\right\rangle\right|^4$ for each  $N_{c,f}$. 
            The suppression of cross sections below the neutrino fog limits the reach of future experiments.
            }
            \label{fig:DD}
\end{figure}

In \cref{fig:DD}, we illustrate the effect of mixings in category II by estimating the direct detection cross sections for benchmark models $(N_c,N_f)=(2,4)$ and $(4,3)$, with $\Lambda_\chi/m_q = 10^{-2}$ and $10^{-6}$. 
We use cross sections computed in Ref.~\cite{Chen:2023bwg} for a pure 5-plet WIMP, scaling the pure 5-plet cross sections by the overlap of the $|5\text{-plet}\ket$ state with the lightest mass eigenstate $|m_{\text{DM}}\ket$. The uncertainty bands in \cref{fig:DD} are inherited from uncertainties in the calculation of Ref.~\cite{Chen:2023bwg} due to \emph{e.g.}~non-perturbative nuclear matrix elements. At sufficiently small masses, the DM candidate may be excluded by existing \cite{LZCollaboration:2024lux} or projected \cite{XENON:2015gkh} constraints, or by future experiments probing down to the neutrino fog. 
As expected from \cref{fig:mdm}, for sufficiently high DM masses (\textit{i.e.}~where $\SU(2)_L$ appears unbroken) the cross section drops rapidly as the mixing between the DM and the 5-plet states approaches zero.
For the parameters shown in the figure, the cross sections are suppressed below the neutrino fog for $m_{\text{DM}}\gsim\mathcal{O}(0.1-1)\,\text{TeV}$, requiring new techniques for probing heavier DM masses in future direct detection experiments,  see \textit{e.g.}~Refs.~\cite{Blanco:2021hlm,Blanco:2022pkt}.\footnote{In these benchmark models, we also find that the mass splittings of our lightest baryons are far above the $\sim 100\,\text{keV}$ threshold required for inelastic direct detection signals \cite{Tucker-Smith:2001myb,Alves:2009nf,SpierMoreiraAlves:2010err,Chang:2010en,Kumar:2011iy,Eby:2023wem}.}

The effect of uncertainties in the cross section is most evident in the $(4,3)$ model with $\Lambda_\chi/m_q = 10^{-6}$. 
In this case, \cref{fig:DD} suggests the reach of future experiments that can probe down to the neutrino fog can vary from below $300\,\text{GeV}$ to $\sim 5\,\text{TeV}$, depending on the uncertainties. 
This large range of possibilities warrants a more precise calculation of the cross sections applicable to a wider range of DM masses in future work.

As noted in Ref.~\cite{Asadi:2024bbq}, the dimension-7 polarizability operators of our dark baryons are not forbidden by $\mathcal{H}$-parity and can give rise to  scattering in direct detection experiments, even when DM is a singlet of SU(2)$_L$.
The cross section from these operators suffers from large uncertainties \cite{Ovanesyan:2014fha,Kavanagh:2018xeh} and is at best detectable in the current experiments for DM masses below a few hundred GeV.
Future searches will not improve current bounds owing to the rapid drop in the cross section as a function of DM mass ($\sigma \propto 1/m_{\text{DM}}^6$). 
We have also checked numerically that in the large DM mass limit, the electroweak loop-induced interactions scale as $\sigma \propto 1/m_{\text{DM}}^4$. Therefore, the polarizability interaction is subdominant despite the small mixing.

We expect in general that other DM candidates in category II will be mostly comprised of singlet states with similarly small mixings of states in larger representations as in the two examples above. Therefore, we call the DM candidates in categories I and II Noble Dark Matter. 
\subsection{Indirect Detection and Astrophysical Searches}
\label{subsec:ID}

As for indirect detection, note that in the heavy quark limit, for the DM masses shown in Fig.~\ref{fig:DD}, the symmetric dark baryon abundance is completely depleted during a 1$^{\mathrm{st}}$ order confinement phase transition by the squeezeout mechanism \cite{Asadi:2021pwo,Asadi:2021yml,Asadi:2022vkc}. Thus, when the phase transition is of the $1^{\mathrm{st}}$ order, which  is anticipated to occur for $N_c > 2$ and $\Lambda_\chi/m_q \lesssim 10^{-2}$ \cite{Svetitsky:1982gs,Alexandrou:1998wv,Kaczmarek:1999mm,Lucini:2003zr,Lucini:2005vg,Aoki:2006we,Saito:2011fs}, we would need to augment the model to generate an asymmetric DM abundance today.

Even in the asymmetric limit, DM can accumulate in the center of stars.
This can occur if the DM has a large enough interaction with SM such that it can lose its kinetic energy and slow down to below the escape velocity of the object.
This accumulation can result in an implosion of the star via formation of a dense core that eventually collapses into a black hole~\cite{Goldman:1989nd,Gould:1989gw,Kouvaris:2010vv,deLavallaz:2010wp,McDermott:2011jp,Kouvaris:2011fi,Guver:2012ba,Kouvaris:2013kra,Bramante:2014zca,Bramante:2017xlb,Kouvaris:2018wnh,Garani:2018kkd,Ellis:2021ztw,Bhattacharya:2023stq}. 
There remains some debate in the literature as to the applicability of these bounds in certain DM ranges or self-interaction rates \cite{Bell:2013xk,Bramante:2013hn,Bramante:2013nma,Gresham:2018rqo}, thus a detailed study of this signal in our model remains an interesting future direction.

It is also argued that in models where there is more than one stable dark particle (whether baryons or nuclei), dark nucleosynthesis processes can give rise to detectable indirect detection signals, even in the asymmetric DM limit \cite{Detmold:2014qqa}. 
While lattice studies have established the existence of  stable nuclei for the case of $N_{c,f}=2$ \cite{Detmold:2014kba}, further studies on other combinations of $N_f$ and $N_c$ would be needed to determine the relevance of dark nucleosynthesis in these theories. 
Such dark nucleosynthesis processes are enhanced if the DM interacts enough with SM to be captured in the core of different celestial bodies.

In the case where a symmetric abundance could arise, such as by going to larger values of $\Lambda_\chi/m_q$ or for $N_c=2$ where the phase transition is of 2$^{\mathrm{nd}}$ order, other indirect detection signals can exist for the model as well. 
For models in category II, there can be direct annihilation of dark baryons to the SM, but the rate is suppressed by the mixing with non-singlet representations as in direct detection. 
The leading indirect detection signal arises from dark baryons annihilating to dark sector glueballs and/or rearranging into mesons. These daughter hadrons may further decay to the SM. 
In particular, as many of these mesons can be long-lived \cite{Bai:2010qg}, this annihilation signal may be detectable in celestial-body focused searches \cite{Leane:2021ihh}. 
A thorough study of all these indirect detection signals, in either the symmetric or asymmetric abundance limit, is left for future work.

\subsection{Collider Searches}
\label{subsec:collider}

Sufficiently high-energy collisions could produce mesons, baryons, or glueballs of the dark sector. Glueballs are the lightest hadrons when $\Lambda_\chi\ll m_q$, but they could only be produced (at one loop) via highly-suppressed dimension-8 operators unless we introduced an additional portal \cite{Juknevich:2009ji,Juknevich:2009gg}. If $\Lambda_\chi\lsim \text{few}\;\text{GeV}$, the $Z$ boson can decay to dark glueballs through a quark loop to three dark gluons. Future colliders may be able to probe this contribution to the invisible width of the $Z$. See Refs.~\cite{FCC:2018byv,FCC:2018evy,Cobal:2020hmk} for invisible $Z$ decay studies at future colliders.

If kinematically accessible, dark mesons and baryons can be produced in colliders via electroweak processes. For example, charged hadrons can be pair-produced, and neutral vector hadrons can mix with the SM photon or $Z$ before decaying \cite{Kilic:2009mi,Kribs:2018ilo}. Mesons (being lighter than baryons for $N_c>2$) would dominate signals, potentially including \emph{e.g.}~missing energy plus initial state radiation \cite{Han:2020uak}, heavy stable charged particles, displaced vertices, disappearing tracks \cite{Mahbubani:2017gjh,Fukuda:2017jmk,Saito:2019rtg,Capdevilla:2021fmj,Capdevilla:2024bwt}, semi-visible jets \cite{Cohen:2015toa,Cohen:2017pzm}, emerging jets \cite{Schwaller:2015gea}, tumblers \cite{Dienes:2021cxr}, and quirks \cite{Kang:2008ea}. For the relevant mass ranges in \cref{fig:DD}, the center-of-mass energy of hard scattering processes in foreseeable colliders would be insufficient to produce the large dark hadron multiplicities required for soft bombs (SUEPs) \cite{Knapen:2016hky,Barron:2021btf}. See Refs.~\cite{Kribs:2016cew,Lee:2018pag,Alimena:2019zri} for reviews.

These signals depend on the lifetimes of mesons, which can span a large range due to an accidental symmetry \cite{Bai:2010qg}. 
If these signals can be tagged efficiently, and if the energy of hard interactions that produce dark quarks and gluons is large enough relative to the dark hadron masses such that a parton shower and jet formation take place within the dark sector, then energy-energy correlators could be used to probe the dark confining phase transition \cite{PhysRevD.102.054012,Lee:2023npz,PhysRevLett.133.071903}.
We leave a comprehensive study of all these collider signals for future work.

\section{Conclusion and Outlook}
\label{sec:conclusion}

We studied a sector with dark quarks transforming under a confining $\SU(N_c)$ as well as an $N_f$-plet of $\SU(2)_L$. 
We described a systematic way to determine dark baryon $\SU(2)_L$ representations using \cref{eq:NfrepNoD} and showed that for various $N_{c}$,$N_{f}$ combinations, there are one or more $\SU(2)_L$ singlets in the lowest-spin baryon spectrum, see \cref{tab:money_table}. 
Using the non-relativistic quark model, we develop a variational method for calculating the baryon mass spectrum in the heavy quark limit, including the electroweak contributions in the inter-quark potential.

We found that the lightest baryons in these models are neutral and mostly overlap with the singlet states. 
For $m_q\gsim \mathcal{O}(1)\;\text{TeV}$, the mixing of the DM candidate with the non-singlet state is significantly suppressed, as illustrated in \cref{fig:mdm}. 
Thus, while the dark sector is in thermal contact with the SM in the early universe, the DM candidate in the hadronic phase has strongly suppressed interactions with SM today.

 leading electromagnetic moments are forbidden by $\mathcal{H}$-parity \cite{Asadi:2024bbq}, and electroweak loop-induced signals are suppressed when the the DM candidates are (approximate) $\SU(2)_L$ singlets. 
It is remarkable that symmetries of the simple UV theory considered in this paper lead to a low energy effective theory of dark matter with such highly suppressed interactions.

We focused on the $(N_c,N_f)=(2,4)$ and $(4,3)$ baryon spectra because they have two neutral states (which are permitted to mix by $\mathcal{H}$-parity), one of which is a singlet. The lowest-spin spectra of other $N_{c,f}$ combinations can contain only a singlet (category I from \cref{tab:money_table}), a singlet with more complicated mixing (category II), or no singlet with a nonetheless viable DM candidate (category III). 
Each of these possibilities warrants further investigation, and we have presented methods to estimate the mass spectra and mixings of states in different $\SU(2)_L$ representations in the heavy quark limit.
In the $m_q<\Lambda_\chi$ regime, the calculation is complicated by relativistic and non-perturbative corrections, but we expect our qualitative conclusions based on perturbative electroweak effects to hold true. In particular, states in each $\SU(2)_L$ representation become approximate mass eigenstates for $m_q\gg m_W$, states in larger representations of $\SU(2)_L$ tend to be heavier, and charged states tend to be heavier than neutral states.

There are various generalizations of the model in \cref{eq:L} that could alter the phenomenology. Adding additional quark generations, scalar quarks, or quarks in different $\SU(N_c)$ representations would complicate the hadron spectrum while preserving $\mathcal{H}$-parity. One could also add additional portals (such as a Higgs portal) to facilitate production and decay of dark hadrons, including glueballs.

It is interesting to compare and contrast our model to theories with elementary WIMPs.  
It is well known that DM composed of the neutral component of a single non-trivial SU(2)$_L$ multiplet (such as Wino) has a large indirect detection signal \cite{Baumgart:2017nsr,Rinchiuso:2018ajn,Baumgart:2018yed,Baumgart:2023pwn} and is widely considered to be ruled out. 
In our composite realization, electroweak multiplets can naturally mix with singlet states.
This implies that considering only individual electroweak multiplets in isolation may prematurely dismiss WIMP-like dark matter. 
Combinations of multiple WIMPs
could have been considered in the context of an EFT of elementary states, but the mass hierarchy and mixings would be free parameters.  We have seen that the IR description of a composite realization of electroweak multiplets provides the mass spectrum and the mixings as calculable outcomes of a simple UV theory.  Therefore, there is a much larger theory space of WIMP-like dark matter with multiple states and associated mixings.

There are of course distinctions between elementary WIMPs and the composite WIMP-like states that we encounter in Nobel Dark Matter. For instance, in addition to the baryon states, there is also a set of meson states that are expected to be lighter than the baryons.  The meson states are anticipated to have important implications for indirect detection (annihilation of dark baryons to dark mesons) as well as collider signals.  We reserve a more thorough analysis of the rich phenomenology for future work.

\section*{Acknowledgments}

We thank Elias Bernreuther, Tom Bouley, Marco Costa, Sam Homiller, Nick Rodd, and Tracy Slatyer for helpful discussions. This work is supported in part by the U.S. Department of Energy under grant number DE-SC0011640.

\clearpage

\appendix

\onecolumngrid

\twocolumngrid
\setcounter{equation}{0}
\setcounter{figure}{0}
\setcounter{table}{0}
\setcounter{section}{0}
\makeatletter
\renewcommand{\theequation}{\thesection\arabic{equation}}
\renewcommand{\thefigure}{\thesection\arabic{figure}}
\renewcommand{\thetable}{\thesection\arabic{table}}

\onecolumngrid

\clearpage

\section{Modifications to the \texorpdfstring{$\SU(2)_L$}{SU(2)L} Running Coupling}

\label{app:running}

The dark quark fields we introduce in \cref{eq:L} modify the $\beta$-function of $\alpha_W$ at scales above $m_q$. At sufficiently large $N_c$ and $N_f$, the appearance of Landau poles necessitates a UV completion at a scale below where the coupling diverges \cite{Djukanovic:2017thn}. At one loop and for renormalization scale $\mu>m_q$ \cite{Alves:2014cda},

\begin{align}
    \frac{1}{\alpha_W(\mu)} &= \frac{1}{\alpha_W(m_q)} - \frac{\beta_0}{2\pi}\ln \frac{\mu}{m_q} \label{eq:aw_running} \\
    & \simeq \frac{1}{\alpha_W(m_Z)} - \frac{\beta_{\text{SM}}}{2\pi}\ln \frac{m_q}{m_Z} - \frac{\beta_0}{2\pi}\ln \frac{\mu}{m_q}\,, \label{eq:aw_running_mz}
\end{align}

\noindent with

\begin{align}
    \beta_0 &= \beta_{\text{SM}} + \frac{4}{3}T_R N_c \\
    \beta_{\text{SM}} &= -\frac{19}{6} \\
    \Tr(J_aJ_b) &= T_R \, \delta_{ab}, \ a,b\in \{x,y,z\} \\
    T_R &= \frac{N_f(N_f^2-1)}{12}\,.
\end{align}

\noindent One obtains \cref{eq:aw_running_mz} from \cref{eq:aw_running} by running $\alpha_W$ from $\mu=m_Z$ to $\mu=m_q$ with the SM field content. 

The $N_f^3$ scaling of $T_R$ quickly spoils the asymptotic freedom of $\SU(2)_L$, as the $\beta$-function is negative only if $N_f=2$ and $N_c\leq 4$. Otherwise, there is a UV Landau pole $\Lambda_W$ where $1/\alpha_W(\Lambda_W) = 0$. Solving \cref{eq:aw_running_mz} for $\Lambda_W$,

\begin{align}
    \Lambda_W &= m_q \left( \frac{m_Z}{m_q} \right)^{\beta_{\text{SM}}/\beta_0} \exp\left( \frac{2\pi}{\beta_0\,\alpha_W(m_Z)} \right) \\
    &= m_q \left( \frac{m_Z}{m_q} \right)^{\frac{-57}{-57+2N_cN_f(N_f^2-1)}} \exp\left( \frac{36\pi}{(-57+2N_cN_f(N_f^2-1))\,\alpha_W(m_Z)} \right)\,.
    \label{eq:LambdaW}
\end{align}

\noindent As shown in \cref{tab:poles}, one quickly encounters sub-Planckian Landau poles ($\Lambda_W < M_{\text{Pl}}\sim 10^{19}\,\text{GeV}$) for modest values of $N_{c,f}$. It was already noted in Ref.~\cite{Asadi:2024bbq} that our model requires a UV completion below the Planck scale (perhaps even below the Grand Unification scale) to facilitate meson decays before Big Bang Nucleosynthesis, but the Landau poles imply that a UV completion is needed at an even lower scale for some $N_{c}$,$N_{f}$. When $\Lambda_W/m_q$ is not very large (say $\lsim 10^3$), one must carefully consider the effects of higher-dimensional operators induced by the UV completion. These effects can include baryon number violation, $\mathcal{H}$-parity violation, and large two-loop corrections to the SM $\SU(3)_{\text{color}}$ and $\text{U}(1)_Y$ couplings. This pathology is particularly pronounced for $N_f\geq 5$, where $\Lambda_W/m_q$ can be $\mathcal{O}(10)$. 

\clearpage

\begin{table}[t]
\begin{minipage}{0.8\linewidth}
    \centering
    \begin{tabular}{c|c|c} \hline
        ($N_c$,~$N_f$) & $\Lambda_W\,(\text{GeV})$ for $m_q=1\,\text{TeV}$ & $\Lambda_W\,(\text{GeV})$ for $m_q=1\,\text{PeV}$ \\
        \hline
         (2,2) & --- & --- \\
         \hline
         (2,3) & $10^{42}$ & $10^{49}$ \\
         \hline
         (2,4) & $10^{11}$ & $10^{15}$ \\
         \hline
         (2,5) & $10^6$ & $10^{10}$ \\
         \hline
         \hline
         (3,2) & --- & --- \\
         \hline
         (3,3) & $10^{20}$ & $10^{25}$ \\
         \hline
         (3,4) & $10^8$ & $10^{11}$ \\
         \hline
         (3,5) & $10^5$ & $10^{8}$ \\
         \hline
         \hline
         (4,2) & --- & --- \\
         \hline
         (4,3) & $10^{14}$ & $10^{18}$ \\
         \hline
         (4,4) & $10^6$ & $10^{10}$ \\
         \hline
         (4,5) & $10^4$ & $10^{8}$ \\
         \hline
         \hline
         (5,2) & $10^{508}$ & $10^{568}$ \\
         \hline
         (5,3) & $10^{11}$ & $10^{15}$ \\
         \hline
         (5,4) & $10^6$ & $10^{9}$ \\
         \hline
         (5,5) & $10^4$ & $10^{7}$ \\ \hline
    \end{tabular}
    \caption{ Order-of-magnitude estimates of $\SU(2)_L$ Landau poles $\Lambda_W$ for various $N_{c,f}$ combinations using \cref{eq:LambdaW}. Dashes indicate where $\SU(2)_L$ remains asymptotically free. Some versions of the theory are safe from sub-Planckian Landau poles, and others require a UV completion at modestly high scales.
    }
    \label{tab:poles}
\end{minipage}
\end{table}

\setcounter{table}{0}

\section{Further Details on Hadron Representations}
\label{app:group_theory}

Our conclusions about the phenomenology of the confining dark sector we present in \cref{sec:model} depend on the possible $\SU(2)_L$ representations of the dark baryons, so we require a systematic approach to enumerate them for any combination of $N_c$ and $N_f$. Future analyses of collider signals and cosmology would also benefit from understanding the meson spectrum and representations. 

\subsection{Spin and Flavor Representations}
\label{subsec:spin_flav_reps}

We employ the quark model \cite{DeRujula:1975qlm,Manohar:1983md,Georgi:1984zwz}, wherein hadron states are products of color, flavor, spin, and spatial wavefunctions. Baryon states with identical fermionic constituents must be totally anti-symmetric under particle exchange. Meson constituents transform in conjugate representations of their symmetry groups, so a multi-particle meson state does not consist of identical particles and therefore need not have definite exchange symmetry. 
Since we are most interested in the lightest hadrons, we restrict our analysis to states with vanishing orbital angular momentum and symmetric spatial wavefunctions, and we emphasize the lowest-spin states. 

Under the assumption of confinement, hadron states transform in the singlet representation of $\SU(N_c)$, which implies baryon color wavefunctions are totally anti-symmetric in exchange of their $N_c$ constituents. With our additional assumption of even spatial wavefunctions, the combined spin-flavor states of the baryons in our model must be totally symmetric in particle exchange. Therefore, baryons with spin-1/2 constituents have spin-flavor states belonging to the symmetric representation of $\SU(2N_f)$ with the Young tableau

\begin{gather} \label{eq:2NfRep}
\begin{ytableau}
       \tikznode{a1}{~} & \none[\cdots] & \tikznode{a2}{~}
\end{ytableau} \\ \notag \\
\text{dimension} = \frac{(2N_f-1+N_c)!}{N_c!\,(2N_f-1)!}\,. \notag
\end{gather}
\tikz[overlay,remember picture]{
\draw[decorate,decoration={brace}] ([yshift=-3mm,xshift=2mm]a2.north east) -- 
([yshift=-3mm,xshift=-2mm]a1.north west) node[midway,below]{$N_c$};
}

\noindent The group is $\SU(2N_f)$ since for each of the $N_f$ possible flavors, there are two possible spin states. The dimension of this representation counts the number of baryon spin-flavor states, including all spin degrees of freedom. However, since we want to understand the hadrons' $\SU(2)_L$ representations, and $\SU(2)_L$ is a subgroup of the $\SU(N_f)$ flavor symmetry (analogous to SM isospin), it is useful to consider the hadron states' flavor representations in $\SU(N_f)$ separately from their spin representations in $\SU(2)_{\text{spin}}$. 

In order for a baryon's combined spin-flavor representation to be totally symmetric, its representations in $\SU(N_f)$ and $\SU(2)_{\text{spin}}$ \textit{must have identical exchange symmetries}. Therefore, those representations must have identical Young tableaux when written out with all $N_c$ boxes. The tableaux for $\SU(2)_{\text{spin}}$ representations have at most two boxes in each column, so the same is true for the relevant $\SU(N_f)$ representations. Thus, spin-$S$ baryon states belong to the flavor representation of $\SU(N_f)$ with the tableau\footnote{A spin-$S$ representation of $\SU(2)_{\text{spin}}$ has a row of $2S$ boxes in its Young tableau after contraction.}

\begin{gather} \label{eq:NfRep}
\begin{ytableau}
       \tikznode{a1}{~} & \none[\cdots] & \tikznode{a2}{~} & \tikznode{a3}{~} & \none[\cdots] & \tikznode{a4}{~} \\
       \tikznode{b1}{~} & \none[\cdots] & \tikznode{b2}{~} & \none & \none & \none
\end{ytableau} \\ \notag \\
\text{dimension} = \frac{(2S+1)(N_f+N_c/2+S-1)!\,(N_f+N_c/2-S-2)!}{(N_f-1)!\,(N_f-2)!\,(N_c/2+S+1)!\,(N_c/2-S)!}\,, \notag
\end{gather}
\tikz[overlay,remember picture]{
\draw[decorate,decoration={brace}] ([yshift=-3mm,xshift=2mm]b2.north east) -- 
([yshift=-3mm,xshift=-2mm]b1.north west) node[midway,below]{$N_c/2-S$};
}
\tikz[overlay,remember picture]{
\draw[decorate,decoration={brace}] ([yshift=-3mm,xshift=2mm]a4.north east) -- 
([yshift=-3mm,xshift=-2mm]a3.north west) node[midway,below]{$2S$};
}

\noindent and the dimension of this representation counts the number of spin-$S$ baryons, see also Ref.~\cite{Mitridate:2017oky} for similar results. Unlike the expression in \cref{eq:2NfRep}, this counting does not include all values of $S_z = -S,\ldots,S$ for the various $|S\,S_z\ket$ baryon spin states. 

As an example of applying \cref{eq:NfRep}, consider the model with $N_c = 5$ and $N_f = 4$. Na\"{i}vely, a candidate flavor representation of $\SU(4)$ for spin-$1/2$ baryons would be 
\begin{equation} \label{eq:candidate_rep}
\ydiagram{2,1,1,1}
\end{equation}

\noindent which can be contracted to the fundamental representation of $\SU(4)$. This may seem to suggest that baryon states can be constructed in the flavor representation in \cref{eq:candidate_rep} and the fundamental representation of $\SU(2)_{\text{spin}}$. This is, however, not true because there is no representation of $\SU(2)_{\text{spin}}$ with the Young tableau in \cref{eq:candidate_rep}. Any spin configuration with $S=1/2$ has different particle exchange symmetries than any flavor configuration in the above representation, so the combined spin-flavor representation would not be totally symmetric. 
The only flavor representation with spin-$1/2$ baryons in this model would be 
\begin{equation} 
\ydiagram{3,2}
\end{equation}

\noindent which is a valid tableau for an $\SU(2)_{\text{spin}}$ representation that can be contracted to the fundamental representation. 
This is a dimension-60 representation of the SU(4) flavor group. 
In \cref{subsec:algorithm}, we describe how to construct baryon spin-flavor wavefunctions using the symmetry of the full $\SU(2N_f)$ representation. We then show in \cref{subsec:more_reps} how to construct the flavor part of the wavefunctions (more efficiently) using the symmetry structure of the flavor representation in \cref{eq:NfRep}. 

\subsection{Finding Hadron Wavefunctions and \texorpdfstring{$\SU(2)_L$}{SU(2)L} Representations}

\label{subsec:algorithm}

Now that we have the flavor representations in $\SU(N_f)$, we can formally understand why the baryon states are in specific $\SU(2)_L$ multiplets. Recall that $\SU(2)_L$ is a subgroup of $\SU(N_f)$. For $\mathcal{H}$-parity symmetric models, we are interested in the embedding of $\SU(2)_L$ in $\SU(N_f)$ where one identifies the $N_f$-dimensional irreducible representation of $\SU(2)_L$ with the fundamental representation of $\SU(N_f)$. One can express the flavor representation in \cref{eq:NfRep} as a direct sum of irreducible $\SU(2)_L$ representations corresponding to this embedding. This is the decomposition shown in Table~\ref{tab:money_table} for several combinations of $N_c$ and $N_f$ and the lowest-spin baryons.

We have an algorithm for finding all hadron spin-flavor wavefunctions with a particular spin state, organizing the states into $\SU(2)_L$ multiplets, and using the wavefunctions to compute the matrix elements needed for the mass calculation. For concreteness, we define a spin-flavor eigenstate as a simultaneous eigenstate of the spin operators $S^2$ and $S_z$ as well as all flavor number operators (\emph{i.e.}~a spin-flavor eigenstate has definite quark flavor content). We likewise define an $\SU(2)_L$ eigenstate as a simultaneous eigenstate of the quadratic Casimir operator $J^2$ and the third generator $J_z$ of $\SU(2)_L$. The eigenvalues of $J_z$ are electric charges $Q$ (as the dark sector particles have vanishing hypercharge), and the eigenvalues of $J^2$ are $J(J+1)$. The spectrum of $J$ quantum numbers indicates the decomposition of the $\SU(N_f)$ representation in \cref{eq:NfRep} into representations of $\SU(2)_L$, and each eigenstate of $J^2$ belongs to a $(2J+1)$-plet of $\SU(2)_L$. Enumerating all of these states requires finding all appropriately symmetrized combinations of flavor and spin configurations, then expressing these in a basis that simultaneously diagonalizes $S^2$, $S_z$, $J^2$, and $J_z$. Our algorithm finds these states for either mesons or baryons, and we specify the differences in those cases below. 

A schematic description of our algorithm is 

\begin{enumerate}
\item Find a complete basis of constituent spin configurations for the desired hadron spin state.

\item Find a complete basis of flavor configurations for the hadrons with the ``middlest" electric charge ($Q=1/2$ when $N_c$ is odd and $N_f$ is even and $Q=0$ otherwise).

\item Combine the spin and flavor eigenstates to obtain a complete basis of spin-flavor eigenstates with this charge. For baryons, symmetrize the states and eliminate linear dependence and overlap of the symmetrized states.

\item Transform the basis of spin-flavor eigenstates to a basis that diagonalizes $J^2$ of $\SU(2)_L$. Deduce the spectrum of hadron $\SU(2)_L$ multiplets from the eigenvalues of $J^2$.

\item Apply the $\SU(2)_L$ ladder operators to the eigenstates of $J^2$ with the middlest charge to find all other $\SU(2)_L$ eigenstates in each multiplet.
\end{enumerate}

\noindent We elaborate on each step below, including examples. We start with the middlest electric charge because all $\SU(2)_L$ multiplets contain a state with this charge. One could alternatively start with the highest charge and repeatedly lower it to construct the multiplets. However, that approach is more complicated because there may be additional states with the lower charge that one would have to find. For example, in the $(N_c,N_f)=(5,3)$ model, if one started with the $Q=3$ state in the baryon 7-plet and lowered it to the $Q=2$ state, one would need to find the $Q=2$ state in the 5-plet, then do the same again for the $Q=1$ states in the 5-plet and triplet. By contrast, by finding all the $Q=0$ states and raising/lowering them to fill out the multiplets, one needs to find all states with a particular charge only once.

One must first specify the desired $|S\,S_z\ket$ hadron spin state. One can then find a complete, orthonormal basis of constituent spin configurations for that state. For mesons with spin-$1/2$ constituents, this basis is the well-known spin singlet or triplet state, so we emphasize the baryon case. One can start with the highest-spin baryon state $\left|S=\frac{N_c}{2}\ S_z=\frac{N_c}{2}\right\ket$ and repeatedly apply the lowering operator $S_-$ until the state has the desired $S_z$. One now has a spin state in the multiplet corresponding to the totally symmetric $\SU(2)_{\text{spin}}$ representation with dimension $N_c+1$. Therefore, the state is composed of all basis state spin configurations with the desired $S_z$. One can use this basis to diagonalize the operator $S^2$ and isolate the eigenstates with the desired $S(S+1)$ eigenvalues. One now has a complete, orthonormal basis of constituent spin configurations with the desired hadron spin state.

As an illustrative example, consider the baryon spectrum in the model with $(N_c,N_f)=(4,3)$. The flavor representations of the required form in \cref{eq:NfRep} are 

\begin{align} \label{eq:43_tableaux}
\ydiagram{2,2}& &\ydiagram{3,1}&  &\ydiagram{4} 
\end{align}

\noindent corresponding to spin-0, spin-1, and spin-2 baryons, respectively. Suppose we desire baryons in the spin state $\left|S=0\ S_z=0\right\ket$. One applies $S_-$ to the highest-spin state $|\!\up\up\up\up\,\ket$ to find the basis of six states with $S_z=0$, each with two spin-up and two spin-down quarks. After diagonalizing $S^2$, one finds two states with $S=0$. A (non-unique) orthonormal basis in which to write these states is

\begin{align}
|s_1\ket &= \frac{1}{\sqrt{12}}\left( 
2|\! \up\up\down\down \,\ket + 2|\! \down\down\up\up \,\ket - |\! \up\down\up\down \,\ket - |\! \down\up\down\up \,\ket - |\! \up\down\down\up \,\ket - |\! \down\up\up\down \,\ket  
\right)\\
|s_2\ket &= \frac{1}{2}\left(
|\! \up\down\up\down \,\ket - |\! \down\up\up\down \,\ket + |\! \down\up\down\up \,\ket - |\! \up\down\down\up \,\ket
\right).
\end{align}

\noindent Note that $|s_1\ket$ is symmetric in exchanges of the first pair and last pair of particles, and $|s_2\ket$ is anti-symmetric in these exchanges. 

Next, one needs a complete, orthonormal basis of flavor configurations with the middlest charge, either $Q=1/2$ or $Q=0$. This can be done similarly to the spin configurations above. Starting with the highest-charge flavor state, one can generate the symmetric $\SU(2)_L$ multiplet by repeatedly applying $J_-$, and the resulting state is a combination of all flavor configurations with the desired charge. Labeling quark flavors by their charge and labeling anti-quark flavors with charge $-Q$ as $\overline Q$, the $SU(2)_L$ generators act on baryon and meson flavor states as 

\begin{align}
J_z \left| Q_1,\ldots,Q_{N_c} \right\rangle =& \ \sum_{i=1}^{N_c} Q_i \ \left| Q_1,\ldots,Q_{N_c} \right\rangle , \\
J_\pm \left| Q_1,\ldots,Q_{N_c} \right\rangle =& \ \sum_{i=1}^{N_c} \sqrt{\frac{N_f^2-1}{4} - Q_i(Q_i\pm 1)} \ \left| Q_1,\ldots, Q_i \pm 1,\ldots,Q_{N_c} \right\rangle , \\
J_z \left| Q_1, \overline Q_2 \right\rangle =& \ (Q_1-Q_2) \left| Q_1, \overline Q_2 \right\rangle , \\
J_\pm \left| Q_1, \overline Q_2 \right\rangle = \ \sqrt{\frac{N_f^2-1}{4} - Q_1(Q_1\pm 1)}& \ \left| Q_1\pm1, \overline Q_2\right\rangle   
 - \sqrt{\frac{N_f^2-1}{4} - Q_2(Q_2\mp 1)} \ \left| Q_1, \overline{Q_2\mp1}\right\rangle .
\end{align}

\noindent We work in the convention with

\begin{align}
J_\pm &= J_x \pm iJ_y\,, \\
J^2 &= J_z^2 + \frac{1}{2} \left( J_+J_- +J_-J_+ \right)\,,
\end{align}

\noindent where $J_x$ and $J_y$ are the first and second generators of $\SU(2)_L$, respectively.

In the example of the $(4,3)$ model, we can label the quark flavors as $+,0,-$, corresponding to positive, zero, and negative electric charges, respectively. The highest-charge state is $|\!++++\ket$. After applying $J_-$, one finds the $Q=0$ flavor configurations 

\begin{align}
|f_1\ket &= | 0000\ket\\
|f_2\ket &= | --++ \ket\\
|f_3\ket &= | - \, 00\,+ \ket,
\end{align}

\noindent plus permutations. Note that $|f_1\ket$ has more exchange symmetries than $|f_2\ket$, which has more than $|f_3\ket$.

Now that one has bases for the spin and flavor configurations, one can find the combined spin-flavor states. For meson states, these are simply the tensor products of each flavor configuration with the spin configuration identified above. However, baryon spin-flavor states must be symmetrized. To find all the needed baryon states, one can start by finding the tensor product state of each flavor configuration with each spin configuration. For each of these tensor product states, written with each particle $p_1,\ldots,p_{N_c}$ having some flavor and spin assignment, one then finds the symmetrized state

\begin{equation} \label{eq:symmetrized_state}
\sum_{\sigma\in \text{perms}(N_c)} \left| p_{\sigma(1)},\ldots, p_{\sigma(N_c)} \right\ket,
\end{equation}

\noindent where $\text{perms}(N_c)$ is the set of all permutations of $N_c$ elements. Any valid baryon spin-flavor eigenstate must be some symmetrized product of a flavor content with a spin configuration, so the set of all states obtained using \cref{eq:symmetrized_state} is an over-complete set from which a complete basis of baryon spin-flavor eigenstates can be extracted. One can iteratively construct an orthogonal basis $\{|b_1\ket,\ldots,|b_n\ket\}$ of states that are linear combinations of non-orthogonal states in the set $\{|v_1\ket,\ldots,|v_n\ket\}$ using the Gram-Schmidt process. One sets $|b_1\ket=|v_1\ket$, then $|b_n\ket$ is constructed using the previously found $n-1$ states as

\begin{equation}
|b_n\ket = |v_n\ket - \sum_{i=1}^{n-1} \frac{\bra b_i | v_n \ket}{\bra b_i | b_i \ket} | b_i \ket.
\end{equation}

\noindent Since the original set of symmetrized baryon states is over-complete, some of the states constructed in this way will have zero norm, and such states can be discarded. One now has a complete, orthogonal basis of meson or baryon spin-flavor eigenstates with the desired spin and the middlest charge. 

In the $(4,3)$ model, symmetrizing the tensor product of the $|f_1\ket$ flavor configuration with either spin configuration results in the the state with zero norm. This is unsurprising, as $|f_1\ket$ is totally symmetric in particle exchange, while neither spin configuration is totally symmetric. This case is analogous to the lack of a spin-$1/2$ baryon with three up quarks in the SM. 

Finding the product of $|f_2\ket$ with either spin state and symmetrizing gives the neutral baryon spin-flavor eigenstate
\begin{equation} \label{eq:B01}
| \mathcal{B}_1^0\ket = \frac{1}{\sqrt{72}} | --++ \ket \otimes \left(
2|\! \up\up\down\down \,\ket + 2|\! \down\down\up\up \,\ket - |\! \up\down\up\down \,\ket - |\! \down\up\down\up \,\ket - |\! \up\down\down\up \,\ket - |\! \down\up\up\down \,\ket
\right) + \text{perms},
\end{equation}

\noindent where ``perms" denotes particle permutations that do not return the same configuration. Doing the same with $|f_3\ket$ results in 

\begin{equation} \label{eq:B02}
| \mathcal{B}_2^0\ket = \frac{1}{12} | -00\,+ \ket \otimes \left(
|\! \up\up\down\down \,\ket + |\! \down\down\up\up \,\ket + |\! \up\down\up\down \,\ket + |\! \down\up\down\up \,\ket - 2 |\! \up\down\down\up \,\ket - 2 |\! \down\up\up\down \,\ket
\right) + \text{perms}.
\end{equation}

\noindent These form a complete basis of neutral baryons with the desired spin, but our algorithm prescribes also finding the symmetrized baryon states generated by the permuted flavor configurations not shown above to ensure all baryons are found. The Gram-Schmidt process eliminates the redundancy from pairing flavor and spin states that provide linearly dependent symmetrized baryon states.

One can determine the spectrum of hadron $\SU(2)_L$ multiplets by using the basis of spin-flavor eigenstates to diagonalize the $J^2$ operator. These multiplets provide the decomposition of the $\SU(N_f)$ flavor representation into irreducible representations of $\SU(2)_L$. Some eigenvalues of $J^2$ may be degenerate (\emph{e.g.}~the scalar singlets for $(N_c,N_f)=(4,5)$), in which case there is not a unique way to write the $\SU(2)_L$ eigenstates as linear combinations of spin-flavor eigenstates. 

Finally, one can repeatedly apply the $J_\pm$ operators to the $\SU(2)_L$ eigenstates with the middlest charge to obtain the spin-flavor wavefunctions for all entries of each multiplet. One now has the full spectrum of lowest-lying hadron spin-flavor states with the desired spin, organized by $\SU(2)_L$ representation and charge. 

For the $(4,3)$ model, using the states in \cref{eq:B01,eq:B02} to diagonalize $J^2$ shows that the linear combination 
\begin{equation} \label{eq:43singlet}
|\mathcal{J}_0^0\ket = - \frac{1}{\sqrt{3}}\, | \mathcal{B}_1^0\ket + \sqrt{\frac{2}{3}}\ | \mathcal{B}_2^0\ket 
\end{equation}

\noindent is an $\SU(2)_L$ singlet, and acting with $J_\pm$ on the orthogonal combination generates the $\SU(2)_L$ 5-plet

\begin{equation} \label{eq:435plet}
\begin{pmatrix}
|\mathcal{J}_2^{++}\ket \\
|\mathcal{J}_2^{+}\ket \\
|\mathcal{J}_2^0\ket \\
|\mathcal{J}_2^{-}\ket \\
|\mathcal{J}_2^{--}\ket
\end{pmatrix}
=
\begin{pmatrix}
| \mathcal{B}^{++}\ket \\
| \mathcal{B}^{+}\ket \\
\sqrt{\frac{2}{3}}\ | \mathcal{B}_1^0\ket + \frac{1}{\sqrt{3}}\, | \mathcal{B}_2^0\ket \\
| \mathcal{B}^{-}\ket \\
| \mathcal{B}^{--}\ket
\end{pmatrix}\,,
\end{equation}

\noindent where

\begin{align} 
| \mathcal{B}^{++}\ket &= \frac{1}{\sqrt{72}} | 00++ \ket \otimes \left(
2|\! \up\up\down\down \,\ket + 2|\! \down\down\up\up \,\ket - |\! \up\down\up\down \,\ket - |\! \down\up\down\up \,\ket - |\! \up\down\down\up \,\ket - |\! \down\up\up\down \,\ket
\right) + \text{perms} \\
| \mathcal{B}^{+}\ket &= \frac{1}{12} | -0++ \ket \otimes \left(
2|\! \up\up\down\down \,\ket + 2|\! \down\down\up\up \,\ket - |\! \up\down\up\down \,\ket - |\! \down\up\down\up \,\ket - |\! \up\down\down\up \,\ket - |\! \down\up\up\down \,\ket
\right) + \text{perms},
\end{align}

\noindent and the negatively-charged states $| \mathcal{B}^{-}\ket$ and $| \mathcal{B}^{--}\ket$ can be found by negating the flavors of the positively-charged states. In the above notation, states labeled with $\mathcal{B}$ are spin-flavor eigenstates. States labeled with $\mathcal{J}$ are $\SU(2)_L$ eigenstates whose subscripts denote the $J$ quantum numbers corresponding to the $J(J+1)$ eigenvalues of $J^2$. The presence of the singlet and 5-plet shows that for the relevant embedding, the $\SU(3)$ flavor representation decomposes to $\SU(2)_L$ representations as 

\begin{align}
\SU(3) &\supset \SU(2)_L \\
\ydiagram{2,2} &\rightarrow \bf{5} \oplus 1 \,, \notag
\end{align}

\noindent as stated in \cref{tab:money_table}.

Our other main example in \cref{subsec:result_spectrum} is the $(N_c,N_f)=(2,4)$ model, for which we briefly walk through the baryon wavefunction derivation here. The $\SU(4)$ flavor representations are

\begin{align}
\ydiagram{1,1} & & \ydiagram{2}
\end{align}

\noindent for spin-0 and spin-1 baryons, respectively. We again specify the $\left|S=0\ S_z=0\right\ket$ spin state, for which the only spin configuration is the well-known anti-symmetric two-body spin-singlet state. The flavor configurations with $Q=0$ are

\begin{align}
\left| -\frac{3}{2}\ +\frac{3}{2}  \right\ket,& & 
\left| +\frac{3}{2}\ -\frac{3}{2}  \right\ket,& &
\left| -\frac{1}{2}\ +\frac{1}{2}  \right\ket,& &
\left| +\frac{1}{2}\ -\frac{1}{2}  \right\ket.
\end{align}

\noindent When paired with the allowed spin configuration and symmetrized, these generate the baryon spin-flavor eigenstates 

\begin{align}
| \text{B}_1^0\ket &= \frac{1}{2}\left( \ \left| -\frac{3}{2}\ +\frac{3}{2} \right\ket 
- \left| +\frac{3}{2}\ -\frac{3}{2} \right\ket 
\right) \otimes  \left(
|\! \up\down \,\ket -|\! \down\up \,\ket 
\right), \\
| \text{B}_2^0\ket &= \frac{1}{2}\left( \ \left| -\frac{1}{2}\ +\frac{1}{2} \right\ket 
- \left| +\frac{1}{2}\ -\frac{1}{2} \right\ket 
\right) \otimes  \left(
|\! \up\down \,\ket -|\! \down\up \,\ket 
\right)\,,
\end{align}

\noindent which, when written in the basis that diagonalizes $J^2$ and acted on with $J_\pm$, provide the $\SU(2)_L$ singlet

\begin{equation} \label{eq:24singlet}
|\text{J}_0^0\ket = \frac{1}{\sqrt{2}} \left( | \text{B}_1^0\ket - | \text{B}_2^0\ket \right)
\end{equation}

\noindent and 5-plet

\begin{equation} \label{eq:245plet}
\begin{pmatrix}
|\text{J}_2^{++}\ket \\
|\text{J}_2^{+}\ket \\
|\text{J}_2^0\ket \\
|\text{J}_2^{-}\ket \\
|\text{J}_2^{--}\ket
\end{pmatrix}
=
\begin{pmatrix}
| \text{B}^{++}\ket \\
| \text{B}^{+}\ket \\
\frac{1}{\sqrt{2}} \left( | \text{B}_1^0\ket + | \text{B}_2^0\ket \right) \\
| \text{B}^{-}\ket \\
| \text{B}^{--}\ket
\end{pmatrix}\,,
\end{equation}

\noindent where

\begin{align}
| \text{B}^{++}\ket &= \frac{1}{2}\left( \ \left| +\frac{1}{2}\ +\frac{3}{2} \right\ket 
- \left| +\frac{3}{2}\ +\frac{1}{2} \right\ket 
\right) \otimes  \left(
|\! \up\down \,\ket -|\! \down\up \,\ket 
\right)
 \\
| \text{B}^{+}\ket &= \frac{1}{2}\left( \ \left| -\frac{1}{2}\ +\frac{3}{2} \right\ket 
- \left| +\frac{3}{2}\ -\frac{1}{2} \right\ket 
\right) \otimes  \left(
|\! \up\down \,\ket -|\! \down\up \,\ket 
\right),
\end{align}

\noindent and so on for the negatively-charged states. These multiplets indicate the decomposition

\begin{align}
\SU(4) &\supset \SU(2)_L \\
\ydiagram{1,1} &\rightarrow \bf{5} \oplus 1 \,, \notag
\end{align}

\noindent for our embedding of $\SU(2)_L$ in flavor $\SU(4)$. 
The only reason why we chose both benchmark models to have the same $\SU(2)_L$ multiplets is because the singlet and 5-plet combination is the simplest case where there is a singlet in the spectrum and mixing of neutral states in different representations.

In the electroweak-symmetric phase of the SM, gauge invariance guarantees that each 5-plet baryon state in \cref{eq:435plet} is a mass eigenstate with exactly the same mass, and likewise for the states in \cref{eq:245plet}. Then, the singlets in \cref{eq:43singlet,eq:24singlet} are mass eigenstates with different masses. In the electroweak-broken phase, charge differences cause the masses of states in the 5-plets to split apart, and the neutral $\SU(2)_L$ eigenstates will no longer be mass eigenstates. \cref{app:mass} details the mass calculation and implications of these observations.

\subsection{More Insights on the \texorpdfstring{$\SU(N_f)\rightarrow\SU(2)_L$}{SU(Nf) to SU(2)L} Decomposition}

\label{subsec:more_reps}

There is more than one way to construct baryon wavefunctions and determine the decomposition of the flavor representation into representations of $\SU(2)_L$. The algorithm we described in \cref{subsec:algorithm} is simple to perform, naturally combines flavor states and spin states, and is independent of our foreknowledge of the flavor representation. The fact that the number of baryons obtained by that algorithm agrees with the dimension of the representation in \cref{eq:NfRep} is thus a valuable consistency check. However, as $N_c$ and $N_f$ become larger, the brute-force symmetrization of states becomes computationally slow (particularly when $N_c\gsim6$). We therefore devised an alternative algorithm to perform the $\SU(N_f)\rightarrow\SU(2)_L$ decomposition by leveraging the symmetry structure of the tableau in \cref{eq:NfRep}.

Our alternative algorithm begins by finding all flavor states with the middlest charge in the same way as described in \cref{subsec:algorithm}. Then, one applies \emph{Young symmetrizers} (see \emph{e.g.}~\cite{Bekaert:2006ix}) to each flavor state.\footnote{One typically applies Young symmetrizers to tensors. We are applying them to components of a tensor.} Consider labeling each box in the tableau in \cref{eq:NfRep} with a number corresponding to each particle. For particle labels in each column of the tableau, one anti-symmetrizes the states over that subset of particles (adding up all permutations of the subset of particles weighted by the parity of the permutation). Then, for particle labels in each row of the tableau, one symmetrizes the states over that subset of particles.\footnote{The choice to anti-symmetrize first and symmetrize second is arbitrary.} Take for example the tableau in \cref{eq:43_tableaux} for spin-0 baryons with $(N_c,N_f)=(4,3)$. One can label the boxes as\footnote{The numbering scheme is arbitrary as one performs the procedure for all relevant permutations.}

\begin{equation} \label{eq:labeled_tableau}
\begin{ytableau}
1 & 2 \\
3 & 4
\end{ytableau}    
\end{equation}

\noindent Then, take the flavor state $|f\ket=|-00+\ket$ as an example. One anti-symmetrizes over particles $(1,3)$ and $(2,4)$

\begin{align}
    |f\ket \overset{[13]}{\rightarrow} |f_{[13]}\ket &= |-00\,+\ket - |00-+\ket \\
    |f_{[13]}\ket \overset{[24]}{\rightarrow}|f_{[13][24]}\ket &= |-00\,+\ket - |00-+\ket - |-+\,00\ket + |0+-\,0\ket\,,
\end{align}

\noindent then symmetrizes over $(1,2)$ and $(3,4)$

\begin{align}
    |f_{[13][24]}\ket \overset{(12)}{\rightarrow} |f_{[13][24](12)}\ket &= |-00\,+\ket - 2\,|00-+\ket - |-+\,00\ket + |0+-\,0\ket \\
    & \ \ \ + |0-0\,+\ket - |+-\,00\ket + |+0-0\ket \\
    |f_{[13][24](12)}\ket \overset{(34)}{\rightarrow} |f_{[13][24](12)(34)}\ket &= |-00\,+\ket - 2\,|00-+\ket - 2\,|-+\,00\ket + |0+-\,0\ket \\
    & \ \ \ + |0-0\,+\ket - 2\,|+-\,00\ket + |+0-0\ket + |-0+0\ket \\
    & \ \ \ -2 \,|00+-\ket + |0+0-\ket + |0-+\,0\ket + |+00-\ket\,.
\end{align}

\noindent Doing this procedure for each middlest-charge flavor state and using the Gram-Schmidt process to eliminate the redundancy of the resulting set of states yields a complete, orthogonal basis of flavor states in the desired flavor representation with the desired charge. One can then diagonalize $J^2$ as above to determine the $\SU(2)_L$ representations. This alternative version of the algorithm makes no reference to spin states, though it could be extended to construct the full spin-flavor wavefunctions.

One can interpret each symmetrizer as a linear operator. Then, the product of symmetrizers corresponding to a Young tableau is itself an operator, which one can call the ``tableau operator". If the symmetries of a particular state are incompatible with the symmetries of the desired flavor representation (\emph{e.g.}~a totally symmetric state when the representation is not totally symmetric), then the tableau operator maps that state to the null state. In other words, the vector space of states whose symmetries are incompatible with the desired representation is the kernel of the tableau operator. By consequence, the dimension of the vector space of states that do have compatible symmetries equals the rank (\emph{i.e.}~the number of linearly independent rows of the matrix) of the tableau operator. Since we are only considering the action of the tableau operator on the space of states with the middlest charge, and each $\SU(2)_L$ multiplet contains exactly one such state, the rank of the tableau operator equals the number of $\SU(2)_L$ multiplets in the decomposition.\footnote{Another (less efficient) way to find the $\SU(2)_L$ decomposition would be to compute the ranks of the tableau operators that act on states with each possible charge. The differences between these ranks indicates the differences in the number of states in the desired flavor representation with each charge, from which one can deduce the number of multiplets with a particular maximum charge. For example, if there are $n$ states with $Q=1$ and $n+2$ states with $Q=0$, then there are two singlets in the spectrum. This way, one can perform the $\SU(2)_L$ decomposition without ever diagonalizing $J^2$.}

With the speed of the Young symmetrizer algorithm, we can easily compute the $\SU(N_f)\rightarrow\SU(2)_L$ decompositions for the larger $N_{c,f}$ shown in \cref{tab:reps_extended}. As discussed in \cref{app:running}, Landau poles quickly appear for the larger representations, so one can consider most of the table a group theory exercise rather than a statement about phenomenology. We observe several interesting patterns in the decompositions, which we describe without proof that they hold for all $N_{c,f}$:

\begin{table}
    \centering
    \begin{tabular}{c|c|c} \hline
        ($N_c$,~$N_f$) & Lowest-Spin Baryon SU(2)$_L$ Representations & Next-to-Lowest-Spin Baryon SU(2)$_L$ Representations \\
        \hline
         (2,2) & $\bm{1}$ & $\bm{3}$ \\
         \hline
         (2,3) & $\bm{3}$ & $\bm{5} \oplus \bm{1}$ \\
         \hline
         (2,4) & $\bm{5} \oplus \bm{1}$ & $\bm{7} \oplus \bm{3}$ \\
         \hline
         (2,5) & $\bm{7} \oplus \bm{3} $ & $\bm{9} \oplus \bm{5} \oplus \bm{1}$ \\
         \hline
         (2,6) & $\bm{9} \oplus \bm{5} \oplus \bm{1}$ & $\bm{11} \oplus \bm{7} \oplus \bm{3}$ \\
         \hline
         \hline
         (3,2) & $\bm{2}$ & $\bm{4}$ \\
         \hline
         (3,3) & $\bm{5} \oplus \bm{3} $ & $\bm{7} \oplus \bm{3}$ \\
         \hline
         (3,4) & $\bm{8} \oplus \bm{6} \oplus \bm{4} \oplus \bm{2}$ & $\bm{10} \oplus \bm{6} \oplus \bm{4}$ \\
         \hline
         (3,5) & $\bm{11} \oplus \bm{9} \oplus \bm{7} \oplus \bm{5} \oplus \bm{5} \oplus \bm{3} $ & $\bm{13} \oplus \bm{9} \oplus \bm{7} \oplus \bm{5} \oplus \bm{1}$ \\
         \hline
         (3,6) & $\bm{14} \oplus \bm{12} \oplus \bm{10} \oplus \bm{8} \oplus \bm{8} \oplus \bm{6} \oplus \bm{6} \oplus \bm{4} \oplus \bm{2}$ & $\bm{16} \oplus \bm{12} \oplus \bm{10} \oplus \bm{8} \oplus \bm{6} \oplus \bm{4}$ \\
         \hline
         \hline
         (4,2) & $\bm{1}$ & $\bm{3}$ \\
         \hline
         (4,3) & $\bm{5} \oplus \bm{1}$ & $\bm{7} \oplus \bm{5} \oplus \bm{3}$ \\
         \hline
         (4,4) & $\bm{9} \oplus \bm{5} \oplus \bm{5} \oplus \bm{1}$ & $\bm{11} \oplus \bm{9} \oplus \bm{7} \oplus \bm{7} \oplus \bm{5} \oplus \bm{3} \oplus \bm{3}$ \\
         \hline
         (4,5) & $\bm{13} \oplus \bm{9} \oplus \bm{9} \oplus \bm{7} \oplus \bm{5} \oplus \bm{5} \oplus \bm{1} \oplus\bm{1}$ & $\bm{15} \oplus \bm{13} \oplus \bm{11} \oplus \bm{11} \oplus \bm{9} \oplus \bm{9} \oplus \bm{7} \oplus \bm{7} \oplus \bm{7} \oplus \bm{5} \oplus \bm{5} \oplus \bm{3} \oplus \bm{3}$ \\
         \hline
         \multirow{2}{*}{(4,6)}
         & $\bm{17} \oplus \bm{13} \oplus \bm{13} \oplus \bm{11} \oplus \bm{9} \oplus \bm{9}$
         & $\bm{19} \oplus \bm{17} \oplus \bm{15} \oplus \bm{15} \oplus \bm{13} \oplus \bm{13} \oplus \bm{11} \oplus \bm{11} \oplus \bm{11} \oplus \bm{11}$
         \\
         &  $\oplus\, \bm{9} \oplus \bm{7} \oplus \bm{5} \oplus \bm{5} \oplus \bm{5} \oplus \bm{1} \oplus \bm{1}$
         & $\oplus\, \bm{9} \oplus \bm{9} \oplus \bm{9} \oplus \bm{7} \oplus \bm{7} \oplus \bm{7} \oplus \bm{7} \oplus \bm{5} \oplus \bm{5} \oplus \bm{3} \oplus \bm{3} \oplus \bm{3}$
         \\
         \hline
         \hline
         (5,2) & $\bm{2}$ & $\bm{4}$ \\
         \hline
         (5,3) & $\bm{7} \oplus \bm{5} \oplus \bm{3} $ & $\bm{9} \oplus \bm{7} \oplus \bm{5} \oplus \bm{3}$ \\
         \hline
         (5,4) & $\bm{12} \oplus \bm{10} \oplus \bm{8} \oplus \bm{8} \oplus \bm{6} \oplus \bm{6} \oplus \bm{4} \oplus \bm{4} \oplus \bm{2}$ & $\bm{14} \oplus \bm{12} \oplus \bm{10} \oplus \bm{10} \oplus \bm{8} \oplus \bm{8} \oplus \bm{6} \oplus \bm{6} \oplus \bm{4} \oplus \bm{4} \oplus \bm{2}$ \\
         \hline
         \multirow{2}{*}{(5,5)} & $\bm{17} \oplus \bm{15} \oplus \bm{13} \oplus \bm{13} \oplus \bm{11} \oplus \bm{11} \oplus \bm{11} \oplus \bm{9} \oplus \bm{9} \oplus \bm{9}$ & 
         $\bm{19} \oplus \bm{17} \oplus \bm{15} \oplus \bm{15} \oplus \bm{13} \oplus \bm{13} \oplus \bm{13} \oplus \bm{11} \oplus \bm{11} \oplus \bm{11} \oplus \bm{9}$ \\
         & $ \oplus \, \bm{9} \oplus \bm{7} \oplus \bm{7} \oplus \bm{7} \oplus \bm{5} \oplus \bm{5} \oplus \bm{5} \oplus \bm{5} \oplus \bm{3} \oplus \bm{3} \oplus \bm{1}$
         & $\oplus\, \bm{9} \oplus \bm{9} \oplus \bm{9} \oplus \bm{7} \oplus \bm{7} \oplus \bm{7} \oplus \bm{7} \oplus \bm{5} \oplus \bm{5} \oplus \bm{5} \oplus \bm{3} \oplus \bm{3} \oplus \bm{1}$\\
         \hline
         \hline
         (6,2) & $\bm{1}$ & $\bm{3}$ \\
         \hline
         (6,3) & $\bm{7} \oplus \bm{3}$ & $\bm{9} \oplus \bm{7} \oplus \bm{5} \oplus \bm{5} \oplus \bm{1}$ \\
         \hline
         \multirowcell{2}{(6,4)} 
         & \multirowcell{2}{$\bm{13} \oplus \bm{9} \oplus \bm{9} \oplus \bm{7} \oplus \bm{5} \oplus \bm{5} \oplus \bm{1} \oplus \bm{1}$} 
         & \multirowcell{2}{$\bm{15} \oplus \bm{13} \oplus \bm{11} \oplus \bm{11} \oplus \bm{11} \oplus \bm{9} \oplus \bm{9} \oplus \bm{7}$ \\
         $\oplus\,\bm{7} \oplus \bm{7} \oplus \bm{7} \oplus \bm{5} \oplus \bm{5} \oplus \bm{3} \oplus \bm{3} \oplus \bm{3}$} \\
         & & \\
         \hline
         \multirowcell{4}{(6,5)} 
         & \multirowcell{4}{$\bm{19} \oplus \bm{15} \oplus \bm{15} \oplus \bm{13} \oplus \bm{13} \oplus \bm{11} \oplus \bm{11} \oplus \bm{11} \oplus \bm{9} \oplus \bm{9}$
         \\
         $\oplus\,\bm{7} \oplus \bm{7} \oplus \bm{7} \oplus \bm{7} \oplus \bm{7} \oplus \bm{5} \oplus \bm{3} \oplus \bm{3} \oplus \bm{3}$} 
         & \multirowcell{4}{$\bm{21} \oplus \bm{19} \oplus \bm{17} \oplus \bm{17} \oplus \bm{17} \oplus \bm{15} \oplus \bm{15} \oplus \bm{15} \oplus \bm{13} \oplus \bm{13} \oplus \bm{13}$ \\
         $\oplus\,\bm{13} \oplus \bm{13} \oplus \bm{13} \oplus \bm{11} \oplus \bm{11} \oplus \bm{11} \oplus\ \bm{11} \oplus \bm{11} \oplus \bm{9} \oplus \bm{9} \oplus \bm{9}$ \\
         $ \oplus\, \bm{9}  \oplus \bm{9} \oplus \bm{9} \oplus \bm{9} \oplus \bm{9} \oplus \bm{7} \oplus \bm{7} \oplus \bm{7} \oplus \bm{7} \oplus \bm{7} \oplus \bm{5} \oplus \bm{5} \oplus \bm{5}$\\
         $ \oplus\, \bm{5}  \oplus \bm{5} \oplus \bm{5} \oplus \bm{5} \oplus \bm{3} \oplus \bm{3} \oplus \bm{1} \oplus \bm{1} \oplus \bm{1}$} 
         \\
         & & \\
         & & \\
         & & \\
         \hline
         \hline
         (7,2) & $\bm{2}$ & $\bm{4}$ \\
         \hline
         (7,3) & $\bm{9} \oplus \bm{7} \oplus \bm{5} \oplus \bm{3}$ & $\bm{11} \oplus \bm{9} \oplus \bm{7} \oplus \bm{7} \oplus \bm{5} \oplus \bm{3}$ \\
          \hline
         \multirowcell{3}{(7,4)} 
         & \multirowcell{3}{$\bm{16} \oplus \bm{14} \oplus \bm{12} \oplus \bm{12} \oplus \bm{10} \oplus \bm{10} \oplus \bm{10}$ \\
         $\oplus\, \bm{8} \oplus \bm{8} \oplus \bm{8} \oplus \bm{6} \oplus \bm{6} \oplus \bm{6} \oplus \bm{4} \oplus \bm{4} \oplus \bm{4} \oplus \bm{2}$} 
         & \multirowcell{3}{$\bm{18} \oplus \bm{16} \oplus \bm{14} \oplus \bm{14} \oplus \bm{14} \oplus \bm{12} \oplus \bm{12} \oplus \bm{12}$ \\
         $\oplus\,\bm{10} \oplus \bm{10} \oplus \bm{10} \oplus \bm{10} \oplus \bm{8} \oplus \bm{8} \oplus \bm{8} \oplus \bm{8}$ \\
         $\oplus\,\bm{6} \oplus \bm{6} \oplus \bm{6} \oplus \bm{6} \oplus \bm{4} \oplus \bm{4} \oplus \bm{4} \oplus \bm{2} \oplus \bm{2}$} \\
         & & \\
         & & \\
         \hline
         \hline
         (8,2) & $\bm{1}$ & $\bm{3}$ \\
         \hline
         (8,3) & $\bm{9} \oplus \bm{5} \oplus \bm{1}$ 
         & $\bm{11} \oplus \bm{9} \oplus \bm{7} \oplus \bm{7} \oplus \bm{5} \oplus \bm{3}$ \\
         \hline
         \multirowcell{3}{(8,4)} 
         & \multirowcell{3}{$\bm{17} \oplus \bm{13} \oplus \bm{13} \oplus \bm{11} \oplus \bm{9} \oplus \bm{9} \oplus \bm{9}$ \\
         $\oplus\, \bm{7} \oplus \bm{5} \oplus \bm{5} \oplus \bm{5} \oplus \bm{1} \oplus \bm{1}$} 
         & \multirowcell{3}{$\bm{19} \oplus \bm{17} \oplus \bm{15} \oplus \bm{15} \oplus \bm{15} \oplus \bm{13} \oplus \bm{13} \oplus \bm{13} \oplus \bm{11} \oplus \bm{11}$ \\
         $\oplus\,\bm{11}  \oplus \bm{11} \oplus \bm{11} \oplus \bm{9} \oplus \bm{9} \oplus \bm{9} \oplus \bm{9} \oplus \bm{7} \oplus \bm{7} \oplus \bm{7}$ \\
         $\oplus\,\bm{7}  \oplus \bm{7} \oplus \bm{7} \oplus \bm{5} \oplus \bm{5} \oplus \bm{5} \oplus \bm{3} \oplus \bm{3} \oplus \bm{3} \oplus \bm{3}$} \\
         & & \\
         & & \\ \hline
    \end{tabular}
    \caption{
    Decomposition of the flavor representation in \cref{eq:NfRep} into representations of its gauged $\SU(2)_L$ subgroup for many $N_{c,f}$ combinations. The lowest-spin baryons are spin-0 for $N_c$ even and spin-1/2 for $N_c$ odd. The next-to-lowest-spin baryons are spin-1 for $N_c$ even and spin-3/2 for $N_c$ odd.
    }
    \label{tab:reps_extended}
\end{table}

\begin{enumerate}
    \item For $N_c=2$, the spin-1 representations for a particular $N_f$ become the spin-0 representations for $N_f+1$.

    \item For $N_c$ even and not divisible by 4, the smallest spin-0 representation is a singlet for $N_f$ even and a triplet for $N_f$ odd. For spin-1, this is reversed.

    \item For $N_c$ divisible by 4, the smallest spin-0 representation is a singlet, and the smallest spin-1 representation is a triplet.

    \item For $N_c=3$, we have found no instances of spin-1/2 singlets. There are, however, instances of spin-3/2 singlets for $N_c=3$ and spin-1/2 singlets for larger odd $N_c$.

    \item When holding $N_c$ fixed and increasing $N_f$ by 1, the size of the largest representation increases by $N_c$.
\end{enumerate}

\noindent We invite the reader to attempt to prove these patterns or observe others.

\clearpage

\subsection{Scalar Quarks}
\label{subsec:squarks}

We have so far considered only spin-1/2 dark quarks, but scalar quarks also have interesting phenomenology. The dark sector Lagrangian would be modified to include the scalar's quartic self-interactions and dimension-4 couplings between the scalar and the SM Higgs boson. The Higgs coupling induces a direct detection signal due to Higgs boson exchange. When the scalar is heavy and integrated out, the Higgs coupling also induces (at one loop) a dimension-6 effective interaction between the Higgs (including its Goldstones) and dark gluons, which facilitates production and decay of dark sector glueballs in a collider context \cite{Juknevich:2009gg,Juknevich:2009ji,Batz:2023zef}. One could still use the quark model in the heavy quark limit, but the inter-quark potential would include terms due to Higgs boson exchange and the quartic coupling in addition to vector boson exchange. 

We can use similar arguments as above to characterize baryon states and their representations. With their bosonic constituents, baryon composites of scalar quarks have totally symmetric wavefunctions. The color wavefunction is still anti-symmetric, and the spins are trivial, so the combination of flavor and spatial wavefunctions must be totally anti-symmetric. Suppose the spatial wavefunction is totally symmetric. Then, the only allowed flavor representation is 

\begin{gather} \label{eq:scalar_anti_sym}
\begin{ytableau}
       \tikznode{a1}{~}  \\
       \none[\raisebox{-2pt}{$\vdots$}] \\
       \tikznode{b1}{~}
\end{ytableau} \\
\text{dimension} = \frac{N_f!}{N_c!(N_f-N_c)!}\,, \notag
\end{gather}
\tikz[overlay,remember picture]{
\draw[decorate,decoration={brace}] ([yshift=-1mm,xshift=-3mm]b1.south west) -- 
([yshift=3mm,xshift=-3mm]a1.north west) node[midway,left]{$N_c$};
}

\noindent since it must be totally anti-symmetric. 

However, this representation of $\SU(N_f)$ does not exist if $N_c>N_f$, as a Young tableau for $\SU(N_f)$ cannot have a column with more than $N_f$ boxes. In such cases, baryons with symmetric spatial wavefunctions are \emph{forbidden}, and the lowest-lying states are $p$-wave. With an anti-symmetric spatial wavefunction, the only allowed flavor representation is 

\begin{gather} \label{eq:scalar_sym}
\begin{ytableau}
       \tikznode{a1}{~} & \none[\cdots] & \tikznode{a2}{~}
\end{ytableau} \\ \notag \\
\text{dimension} = \frac{(N_f-1+N_c)!}{N_c!\,(N_f-1)!}\,. \notag
\end{gather}
\tikz[overlay,remember picture]{
\draw[decorate,decoration={brace}] ([yshift=-3mm,xshift=2mm]a2.north east) -- 
([yshift=-3mm,xshift=-2mm]a1.north west) node[midway,below]{$N_c$};
}

\noindent These representations can be decomposed into representations of $\SU(2)_L$, as shown in \cref{tab:squarks}. Note in particular that if $N_c=N_f$, the representation in \cref{eq:scalar_anti_sym} is the trivial representation of $\SU(N_f)$, so the only possible $\SU(2)_L$ multiplet is the singlet. The version of our algorithm with Young symmetrizers naturally performs the decomposition for baryons with scalar constituents. For the $\SU(N_f)$ representations we have decomposed, singlets appear slightly less often compared to the case of spin-1/2 constituents.

We can sort each spectrum of multiplets into categories analogous to those defined in \cref{subsec:IR}:

\begin{enumerate}[label=\Roman*.]

    \item There is only a singlet ($N_c=N_f$).

    \item There is a singlet accompanied by larger odd representations.

    \item There are no singlets, but there are neutral states.

    \item There are no neutral states.

\end{enumerate}

\noindent Note that the decomposition of the flavor representation into representations of SU(2)$_L$ for $N_c=2$ is the same for lowest-lying baryons with scalar constituents and spin-0 baryons with spin-1/2 constituents because in this case, the representations in \cref{eq:NfRep} and \cref{eq:scalar_anti_sym} are equivalent.

\clearpage

\begin{table}[t]
    \centering
    \begin{minipage}{0.8\linewidth}
    \begin{tabular}{c|c|c}
    \hline
        ($N_c$,~$N_f$) & Baryon SU(2)$_L$ Representations (Scalar Constituents) & Category  \\
        \hline
         (2,2) & $\bm{1}$ & I  \\
         \hline
         (2,3) & $\bm{3}$ & III \\
         \hline
         (2,4) & $\bm{5} \oplus \bm{1}$ & II  \\
         \hline
         (2,5) & $\bm{7} \oplus \bm{3} $ & III \\
         \hline
         (2,6) & $\bm{9} \oplus \bm{5} \oplus \bm{1}$ & II \\
         \hline
         \hline
         (3,2) & $\bm{4}$ & IV \\
         \hline
         (3,3) & $\bm{1} $ & I \\
         \hline
         (3,4) & $\bm{4}$ & IV  \\
         \hline
         (3,5) & $\bm{7}  \oplus \bm{3} $ & III \\
         \hline
         (3,6) & $\bm{10} \oplus \bm{6} \oplus \bm{4}$ & IV  \\
         \hline
         \hline
         (4,2) & $\bm{5}$ & III \\
         \hline
         (4,3) & $\bm{9} \oplus \bm{5} \oplus \bm{1}$ & II \\
         \hline
         (4,4) & $\bm{1}$ & I \\
         \hline
         (4,5) & $\bm{5}$ & III \\
         \hline
         \multirow{1}{*}{(4,6)}
         & $\bm{9} \oplus \bm{5} \oplus \bm{1}$  & II
        \\
         \hline
         \hline
         (5,2) & $\bm{6}$ & IV  \\
         \hline
         (5,3) & $\bm{11} \oplus \bm{7} \oplus \bm{3} $ & III  \\
         \hline
         (5,4) & $\bm{16} \oplus \bm{12} \oplus \bm{10} \oplus \bm{8} \oplus \bm{6} \oplus \bm{4}$  & IV \\
         \hline
         \multirow{1}{*}{(5,5)} & $\bm{1}$ & I \\
         \hline
         \multirow{1}{*}{(5,6)} & $\bm{6}$ & IV \\
         \hline
         \hline
         (6,2) & $\bm{7}$ & III  \\
         \hline
         (6,3) & $\bm{13} \oplus \bm{9} \oplus \bm{5} \oplus \bm{1}$ & II \\
         \hline
         \multirowcell{1}{(6,4)} 
         & \multirowcell{1}{$\bm{19} \oplus \bm{15} \oplus \bm{13} \oplus \bm{11} \oplus \bm{9} \oplus \bm{7} \oplus \bm{7} \oplus \bm{3} $ 
         }  & III \\
         \hline
         \multirowcell{2}{(6,5)} 
         & \multirowcell{2}{$\bm{25} \oplus \bm{21} \oplus \bm{19} \oplus \bm{17} \oplus \bm{17} \oplus \bm{15} \oplus \bm{13} \oplus \bm{13} \oplus \bm{13}$  
         \\
         $\oplus\,\bm{11} \oplus \bm{9} \oplus \bm{9} \oplus \bm{9} \oplus \bm{7} \oplus \bm{5} \oplus \bm{5} \oplus \bm{1} \oplus \bm{1}$} & \multirowcell{2}{II} \\
         & \\
         \hline
         (6,6) & $\bm{1} $ & I \\
         \hline
         \hline
         (7,2) & $\bm{8}$ & IV \\
         \hline
         (7,3) & $\bm{15} \oplus \bm{11} \oplus \bm{7} \oplus \bm{3}$ & III  \\
         \hline
         (7,4) & $\bm{22} \oplus \bm{18} \oplus \bm{16} \oplus \bm{14} \oplus \bm{12} \oplus \bm{10} \oplus \bm{10} \oplus \bm{8} \oplus \bm{6} \oplus \bm{4}$ & IV  \\
         \hline
         \hline
         (8,2) & $\bm{9}$ & III \\
         \hline
         (8,3) &  $\bm{17} \oplus \bm{13} \oplus \bm{9} \oplus \bm{5} \oplus \bm{1}$  & II \\
         \hline
         \multirowcell{2}{(8,4)} &  \multirowcell{2}{$\bm{25} \oplus \bm{21} \oplus \bm{19} \oplus \bm{17} \oplus \bm{15} \oplus \bm{13} $ \\
         $\oplus\, \bm{13} \oplus \bm{11} \oplus \bm{9} \oplus \bm{9} \oplus \bm{7} \oplus \bm{5} \oplus \bm{1}$ 
         }  & \multirowcell{2}{II} \\
         & \\
         \hline
    \end{tabular}
    \caption{
     Decomposition of the flavor representation in \cref{eq:scalar_anti_sym} (for $N_c\leq N_f$) or \cref{eq:scalar_sym} (for $N_c > N_f$) for baryons with scalar constituents, as in \cref{tab:reps_extended}. We also indicate the category of each spectrum of multiplets defined in \cref{subsec:squarks}.}
    \label{tab:squarks}
    \end{minipage}
\end{table}

\clearpage

\section{Further Details on the Mass Calculation}
\label{app:mass}

In the limit where the dark quark mass $m_q$ is significantly larger than the dark sector confinement scale $\Lambda_\chi$, mesons and baryons are well-approximated as weakly-coupled bound states of non-relativistic constituents. In Ref.~\cite{DeRujula:1975qlm}, this quark model picture is applied to the SM, where the physical quark masses are less than the confinement scale, and non-perturbative effects give large contributions to hadron masses. Nonetheless, this approach predicts SM hadron mass splittings fairly well using non-perturbative dressed quark masses. In the heavy quark limit of our model, we treat non-perturbative and relativistic effects as negligible, so we expect the quark model prediction to be a valid approximation of hadron masses (not only their splittings) using physical quark masses. We cannot reliably compute the hadron masses when $m_q$ is not significantly greater than $\Lambda_\chi$, but our qualitative conclusions concerning relative mass splittings and mixings likely still hold as these are driven by perturbative electroweak effects.

\subsection{The Mass Hamiltonian}
\label{app:Hamiltonian}

The mass operator in \cref{eq:M} comes from the Hamiltonian

\begin{equation}
H = H_{\text{free}} + \sum_{i<j}V_{ij}\,,
\end{equation}

\noindent where $V_{ij}$ is the inter-quark potential, and  $H_{\text{free}}$ is the free Hamiltonian 

\begin{equation}
H_{\text{free}}=\sum_{i} \left(
m_i + \frac{p_i^2}{2m_i}
\right),
\end{equation}

\noindent in the non-relativistic approximation. At first order in the gauge couplings, the inter-quark potential comes from exchange of a single gauge boson. For massless gauge bosons (\emph{i.e.}~the photon and dark gluons) and at first order in the non-relativistic expansion in $p^2/m_q^2$, this potential is the Fermi-Breit potential familiar from the Hydrogen atom in non-relativistic quantum mechanics. 

In addition to the dark sector coupling, the dark gluon potential in \cref{eq:M} includes a color factor $-(N_c+1)/(2N_c)$ from contractions of $\SU(N_c)$ generators in the fundamental representation with anti-symmetric baryon color wavefunctions. For mesons, this factor should be replaced with $-(N_c^2-1)/(2N_c)$, where the sign is from fermion statistics, and the $N_c$-dependence is from contracting $\SU(N_c)$ generators with color-singlet wavefunctions. In the SM with $N_c=3$, these color factors reproduce the $-2/3$ for baryons and $-4/3$ for mesons found in Ref.~\cite{DeRujula:1975qlm}. 

In the electroweak-symmetric limit, the photon and $Z$ contributions in \cref{eq:M} sum to the contribution of the third $\SU(2)_L$ gauge boson, the $W^3$, since $\alpha_{\text{EM}}+\alpha_W\cos^2\theta_W = \alpha_W$. The raising and lowering operators in \cref{eq:M} allow the $W$ boson to mix hadron states with different flavor content, which is why the spin-flavor eigenstates described in \cref{app:group_theory} are not in general equivalent to $\SU(2)_L$ eigenstates or mass eigenstates. In the $m_q,\Lambda_\chi \gg m_W$ limit, electroweak symmetry breaking is ``small" in the dark sector, and mass eigenstates should be nearly identical to $\SU(2)_L$ eigenstates as shown in \cref{fig:mdm}. In the $m_q,\Lambda_\chi \ll m_W$ limit where one can integrate out the $W$, spin-flavor eigenstates should be nearly identical to mass eigenstates as their mixing due to the $W$ is suppressed. We indeed see this behavior in \cref{fig:mdm}, as the overlaps of the mass eigenstates with the $\SU(2)_L$ eigenstates approach the overlaps of the spin-flavor eigenstates with the $\SU(2)_L$ eigenstates as $m_q$ becomes small. 

In order to incorporate finite-mass effects of the $W$ and $Z$, we require a generalization of the Fermi-Breit potential whose $0^\text{th}$ order term in the non-relativistic expansion is the Yukawa potential rather than the Coulomb potential. Ref.~\cite{DeSanctis:2009zz} derived that, up to color factors and couplings, the inter-quark potential is 

\begin{align} \label{eq:FermiBreit_gen}
V_{ij} =& V_0(r) \\
& + \left( \frac{1}{8m_i^2} + \frac{1}{8m_j^2} \right) \nabla^2V_0(r) \label{eq:Darwin} \\
&+\left[ 
\left( \frac{1}{2m_i^2} + \frac{1}{m_im_j}\right) {\bm S}_i
+ \left( \frac{1}{2m_j^2} + \frac{1}{m_im_j}\right) {\bm S}_j
\right] \cdot {\bm{\ell}}\,\frac{V_0^\prime(r)}{r} \label{eq:spin_orbit} \\
&+ \frac{1}{m_im_j} \left[
\frac{2}{3} {\bm S}_i\cdot {\bm S}_j \nabla^2V_0(r) \right. \label{eq:spin_spin} \\
 &+\left.\left(
({\bm S}_i\cdot\hat{r})({\bm S}_j\cdot\hat{r})-\frac{1}{3}{\bm S}_i\cdot {\bm S}_j
\right)
\left(
V_0^{\prime\prime}(r)-\frac{V_0^\prime(r)}{r}
\right)
\right] \label{eq:tensor_force} \\
&-\frac{1}{8m_im_j}\left(
\left\{ {\bm p_i}\cdot {\bm p_j}, V_0(r) \right\}+2 p_i^\alpha V_0(r) p_j^\alpha
\right) \label{eq:orbit_orbit1} \\
&-\frac{1}{4m_im_j} \left( 
\left\{ p_i^\alpha p_j^\beta , \hat{r}^\alpha \hat{r}^\beta W(r)\right\}
+ 2 p_i^\alpha  \hat{r}^\alpha \hat{r}^\beta W(r)p_j^\beta
\right)\, \label{eq:orbit_orbit2}
\end{align}

\noindent at $\mathcal{O}(p^2/m^2)$, where $V_0(r)$ is the potential at $0^\text{th}$ order in the non-relativistic expansion, $W(r)=-rV_0^\prime(r)/2$, ${\bm{\ell}}$ is the angular momentum operator, $\hat{r}$ is the unit vector in the direction of the inter-quark separation ${\bm{r}} = {\bm{r}}_i - {\bm{r}}_j$, and $\alpha,\beta$ are Cartesian indices.\footnote{In standard derivations of the Fermi-Breit potential (\emph{e.g.}~Refs.~\cite{berestetskii1982quantum,Donoghue:1992dd}), the ordering of position and momentum operators is not handled carefully, so applying the same calculations to massive boson exchange does not ensure a Hermitian potential. When transforming the potential from momentum space to position space, one must keep initial and final momentum operators written on opposite sides of position operators so that $p_{\text{initial}}$ acts on $|p_{\text{initial}} \ket$ and $p_{\text{final}}$ acts on $\bra p_{\text{final}}|$.} One may recognize \cref{eq:Darwin,eq:spin_orbit} as the fine-structure correction and \cref{eq:spin_spin,eq:tensor_force} as the hyperfine-structure correction.

For hadrons with symmetric spatial wavefunctions, the spin-orbit coupling in \cref{eq:spin_orbit} and tensor force contribution in \cref{eq:tensor_force} do not contribute, since $\bm{\ell}$ has odd parity and $\langle(\hat{r}\cdot{\bm S}_i)(\hat{r}\cdot{\bm S}_j)\rangle \rightarrow \langle{\bm S}_i \cdot {\bm S}_j \rangle/ 3$ \cite{Griffiths_Schroeter_2018}. The remaining contributions are $V_0(r)$, the Darwin term in \cref{eq:Darwin}, the spin-spin coupling in \cref{eq:spin_spin}, and the orbit-orbit coupling in \cref{eq:orbit_orbit1,eq:orbit_orbit2}. Using $V_0(r) = e^{-m_Vr}/r$ for the Yukawa potential, expanding out the momentum operators in terms of gradients, taking the expectation value with respect to a symmetric spatial wavefunction, setting $m_{i,j}=m_q$, and re-inserting coupling coefficients, one recovers \cref{eq:M}.

The contact terms that arise from the spin-spin coupling and Darwin terms are not suppressed by the mass of the exchanged boson. It is unclear how important the contact terms from $W$ and $Z$ exchange are to the quark model calculation in the SM, which has additional complications of chiral couplings (which \cref{eq:FermiBreit_gen} does not accommodate) and non-trivial relativistic and non-perturbative corrections. In our model, these contact terms are indispensable for correctly predicting which baryons species is lightest, particularly when $m_W\ll m_q$.

\subsection{Quark Mass Splittings}
\label{app:quark_splittings}

Invariance of the Lagrangian under $\SU(2)_L$ requires that the tree-level quark masses in our model must be degenerate. Spontaneous breaking of electroweak symmetry breaks that degeneracy at one loop due to self-energy corrections. These have the form

\begin{align} \label{eq:SE_diagram}
\feynmandiagram[horizontal=i1 to a,layered layout] {
i1 [particle=\(\bf{Q}\)] -- [fermion] a -- [fermion] b -- [fermion] f1 [particle=\(\bf{Q}\)],
a -- [photon, half left] b 
};
\end{align}

\noindent where the boson in the loop can be a photon, dark gluon, $W$, or $Z$. In the mass Hamiltonian, we work to leading order in the gauge couplings, so corrections to $m_q$ in the inter-quark potential are higher-order. In the free Hamiltonian, however, the one-loop correction to $m_q$ is at the same order in perturbation theory as the inter-quark potential. We therefore use

\begin{equation}\label{eq:mphys}
m_{i} = m_q - \Sigma_i^{1\text{-loop}}(m_q) + \mathcal{O}(\alpha^2)\,,
\end{equation}

\noindent where $\Sigma_i^{1\text{-loop}}(m_q)$ is the one-loop self-energy of the $i^{\text{th}}$ quark evaluated at $\slashed{p}=m_q$, and $\mathcal{O}(\alpha^2)$ represents corrections at higher order in the gauge couplings \cite{Schwartz:2014sze}. Then, the free Hamiltonian has the form

\begin{equation}
H_{\text{free}}=\sum_{i} \left(
m_q + \frac{p_i^2}{2m_q} + \left(
\frac{p_i^2}{2m_q^2} - 1
\right)
\Sigma_i^{1\text{-loop}}(m_q) 
\right)
+ \mathcal{O}(\alpha^2)\,.
\end{equation}

\noindent This correction can cause our predictions for baryon masses to be greater than $N_c m_q$, even when the expectation value of the inter-quark potential is negative. 

The self-energy is given by 

\begin{align} \label{eq:self_energy}
\Sigma_i^{1\text{-loop}}(m_q) =& -\frac{m_q}{2\pi}\left[ \vphantom{\frac{N^2}{N}} \left(\alpha_{\text{EM}}Q_i^2 + \frac{N_c^2-1}{2N_c}\alpha_\chi\right)\,f(0) + \alpha_W\cos^2(\theta_W)Q_i^2\,f(m_Z/m_q) \right. \\
&+\left. \alpha_W\left( \frac{N_f^2 - 1}{4} - Q_i^2 \right) f(m_W/m_q) \vphantom{\frac{N}{N}} \right], \notag
\end{align}

\noindent where

\begin{equation}
f(r)=\frac{1}{2} \left(4+3 \ln \left(\frac{\mu ^2}{m_q^2}\right)-r^4 \ln r+r^2-\frac{r}{2} \sqrt{r^2-4} \left(r^2+2\right) \ln \left(\frac{r^2-2-r\sqrt{r^2-4}}{2} \right)\right),
\end{equation}

\noindent with $\mu$ the renormalization scale, using Feynman gauge and the $\overline{\text{MS}}$ scheme.\footnote{For the loops with massless gauge bosons, an apparent infrared divergence can be regulated by introducing a gauge boson mass and taking the limit of that mass going to zero. Despite the logarithms, $f(r)$ has a well-defined $r\to 0^+$ limit.} We define the tree-level mass to equal $m_q$ at the scale $m_q$ and therefore evaluate the self-energy and running couplings therein at $\mu=m_q$. One should interpret the $Q_i^2$ factors as operators acting on spin-flavor wavefunctions. Note that in the electroweak-symmetric limit, the terms in \cref{eq:self_energy} with $Q_i^2$ cancel, and the remaining $(N_f^2-1)/4$ prefactor on the electroweak contribution is the Casimir invariant $C_2$ defined by $J^a_{ik}J^a_{kj}=C_2\delta_{ij}$ for the $N_f$-dimensional representation of $\SU(2)_L$. This factor is the $\SU(2)_L$ analog of the $(N_c^2-1)/(2N_c)$ color factor that appears in the dark gluon contribution \cite{Peskin:1995ev}. 

So far, our discussion of the mass calculation has been in the effective field theory of quarks rather than the effective field theory of hadrons. In the theory of hadrons, mass splittings derive from diagrams of the same form as \cref{eq:SE_diagram} with the the quark line replaced by hadron and the gauge boson a photon, $W$, or $Z$. In the limit where $W$ and $Z$ cannot resolve the structure of the hadron, \emph{i.e.}~when $1/m_{W,Z}$ is much greater than the hadron's characteristic size scale, we expect the quark model calculation of hadron mass splittings to approximately match onto the mass splittings computed in the effective field theory of hadrons. As computed in Ref.~\cite{Cirelli:2005uq}, for vector-like fermions or scalars much heavier than $m_{W,Z}$ with zero hypercharge in some multiplet of $\SU(2)_L$, the mass splitting between the $Q=1$ and $Q=0$ components of the multiplet is 

\begin{equation} \label{eq:charge_neutral_splitting}
M_{Q=1}-M_{Q=0} \simeq \alpha_Wm_W\sin(\theta_W/2)\simeq 166\,\text{MeV}.
\end{equation}

\noindent In the limit $m_q,\Lambda_\chi\gg m_W$, our prediction for the splitting between singly-charged and neutral hadrons in the same $\SU(2)_L$ multiplet approximately reproduces this result. This serves as a check on our calculation.

\subsection{Spin-flavor Matrix Elements}
\label{subsec:sfmatrix}

The charge, spin, and ladder operators in \cref{eq:M} act on the spin-flavor wavefunctions found in \cref{subsec:algorithm}. Here, we discuss the actions of these operators on $\SU(2)_L$ eigenstates in the $(N_c,N_f)=(4,3)$ and $(2,4)$ models. We use the following notation for these spin-flavor operators as matrices acting on baryon states:

\begin{equation}
\bra \mathcal{J}^0 | Q_iQ_j | \mathcal{J}^0 \ket \equiv
\begin{pmatrix}
\bra \mathcal{J}^0_0 | \\
\bra \mathcal{J}^0_2 |
\end{pmatrix}
\begin{pmatrix}
\sum\limits_{i<j}Q_iQ_j
\end{pmatrix}
\begin{pmatrix}
| \mathcal{J}^0_0 \ket &
| \mathcal{J}^0_2 \ket
\end{pmatrix},
\end{equation}

\noindent and so on for other operators and $\SU(2)_L$ eigenstates. We abbreviate $(J_{+}^iJ_{-}^j+J_{-}^iJ_{+}^j)/2$ as $J_{\pm}^iJ_{\mp}^j$. As discussed in \cref{app:quark_splittings}, we will need the $\sum_i Q_i^2$ operator in addition to the operators appearing in the inter-quark potential. When acting on hadron states that are not spin-flavor eigenstates, the operators $Q_iQ_j$, $Q_iQ_j \bm{S}_i\cdot\bm{S}_j$, and $Q_i^2$ have off-diagonal matrix elements.

In the $(4,3)$ model, the required matrix elements are 

\begin{align}
\bra \mathcal{J}^0 | Q_iQ_j | \mathcal{J}^0 \ket &= 
\begin{pmatrix}
-4/3 & \sqrt{2}/3 \\
\sqrt{2}/3 & -5/3
\end{pmatrix}
&  
\bra \mathcal{J}^0 | Q_iQ_j \bm{S}_i\cdot\bm{S}_j| \mathcal{J}^0 \ket &= 
\begin{pmatrix}
2/3 & -11/\sqrt{72} \\
-11/\sqrt{72} & 19/12
\end{pmatrix}
\label{eq:QQ43} \\ 
\bra \mathcal{J}^0 | J_{\pm}^iJ_{\mp}^j | \mathcal{J}^0 \ket &= 
\begin{pmatrix}
-8/3 & -\sqrt{2}/3 \\
-\sqrt{2}/3 & 2/3
\end{pmatrix}
&  
\bra \mathcal{J}^0 | J_{\pm}^iJ_{\mp}^j \bm{S}_i\cdot\bm{S}_j| 
\mathcal{J}^0 \ket &= 
\begin{pmatrix}
4/3 & 11/\sqrt{72} \\
11/\sqrt{72} & 7/6
\end{pmatrix}
\label{eq:JJ43} \\
\bra \mathcal{J}^0 | Q_i^2 | \mathcal{J}^0 \ket &=
\begin{pmatrix}
8/3 & -\sqrt{8}/3 \\
-\sqrt{8}/3 & 10/3
\end{pmatrix}
& 
\bra \mathcal{J}^0 | \bm{S}_i\cdot\bm{S}_j| \mathcal{J}^0 \ket &=
\begin{pmatrix}
-3/2 & 0 \\
0 & -3/2
\end{pmatrix} \\
\notag \\
\bra \mathcal{J}^+ | Q_iQ_j | \mathcal{J}^+ \ket &= -1
&
\bra \mathcal{J}^+ | Q_iQ_j \bm{S}_i\cdot\bm{S}_j| \mathcal{J}^+ \ket &=5/4 \\
\bra \mathcal{J}^+ | J_{\pm}^iJ_{\mp}^j | \mathcal{J}^+ \ket &= 0
&
\bra \mathcal{J}^+ | J_{\pm}^iJ_{\mp}^j \bm{S}_i\cdot\bm{S}_j| 
\mathcal{J}^+ \ket &= 3/2 \\
\bra \mathcal{J}^+ | Q_i^2 | \mathcal{J}^+ \ket &=
3
& 
\bra \mathcal{J}^+ | \bm{S}_i\cdot\bm{S}_j| \mathcal{J}^+ \ket &=
-3/2 \\
\notag \\
\bra \mathcal{J}^{++} | Q_iQ_j | \mathcal{J}^{++} \ket &= 1
&
\bra \mathcal{J}^{++} | Q_iQ_j \bm{S}_i\cdot\bm{S}_j| \mathcal{J}^{++} \ket &=1/4 \\
\bra \mathcal{J}^{++} | J_{\pm}^iJ_{\mp}^j | \mathcal{J}^{++} \ket &= -2
&
\bra \mathcal{J}^{++} | J_{\pm}^iJ_{\mp}^j \bm{S}_i\cdot\bm{S}_j| 
\mathcal{J}^{++} \ket &= 5/2 \\
\bra \mathcal{J}^{++} | Q_i^2 | \mathcal{J}^{++} \ket &=
2
& 
\bra \mathcal{J}^{++} | \bm{S}_i\cdot\bm{S}_j| \mathcal{J}^{++} \ket &=
-3/2
\end{align}

\noindent The negatively-charged baryons have the same matrix elements as the corresponding positively-charged baryons. 

One can see that these matrix elements are consistent with gauge invariance in the electroweak-symmetric phase. The charge operators encode the contribution of the third $\SU(2)_L$ force carrier, the $W^3$, and the ladder operators encode the contributions of the $W^1$ and $W^2$. When $m_{W,Z}=0$, the off-diagonal elements in \cref{eq:QQ43,eq:JJ43} cancel between the $W^{1,2}$ and $W^3$ contributions, and this cancellation occurs independently for the operators with and without spin. Therefore, the neutral $\SU(2)_L$ eigenstates are mass eigenstates in this limit, as required. Moreover, observing the second diagonal entries for the neutral baryon matrices (corresponding to the neutral component of the baryon 5-plet) and the matrix elements for the charged baryons, the sum of the $W^{1,2}$ and $W^3$ contributions is identical for each 5-plet baryon. This is again true for the operators with and without spin independently. Therefore, the baryon 5-plet is exactly mass-degenerate in the symmetric phase, as required by gauge invariance. These observations are an important check on the validity of our approach and are true regardless of the spatial expectation value computation.

In the $(2,4)$ model,

\begin{align}
\bra \text{J}^0 | Q_iQ_j | \text{J}^0 \ket &= 
\begin{pmatrix}
-5/4 & 1 \\
1 & -5/4
\end{pmatrix}
&  
\bra \text{J}^0 | Q_iQ_j \bm{S}_i\cdot\bm{S}_j| \text{J}^0 \ket &= 
\begin{pmatrix}
15/16 & -3/4 \\
-3/4 & 15/16
\end{pmatrix} \\ 
\bra \text{J}^0 | J_{\pm}^iJ_{\mp}^j | \text{J}^0 \ket &= 
\begin{pmatrix}
-5/2 & -1 \\
-1 & 1/2
\end{pmatrix}
&  
\bra \text{J}^0 | J_{\pm}^iJ_{\mp}^j \bm{S}_i\cdot\bm{S}_j| 
\text{J}^0 \ket &= 
\begin{pmatrix}
15/8 & 3/4 \\
3/4 & -3/8
\end{pmatrix} \\
\bra \text{J}^0 | Q_i^2 | \text{J}^0 \ket &=
\begin{pmatrix}
5/2 & -2 \\
-2 & 5/2
\end{pmatrix}
& 
\bra \text{J}^0 | \bm{S}_i\cdot\bm{S}_j| \text{J}^0 \ket &=
\begin{pmatrix}
-3/4 & 0 \\
0 & -3/4
\end{pmatrix} \\
\notag \\
\bra \text{J}^+ | Q_iQ_j | \text{J}^+ \ket &= -3/4
&
\bra \text{J}^+ | Q_iQ_j \bm{S}_i\cdot\bm{S}_j| \text{J}^+ \ket &=9/16 \\
\bra \text{J}^+ | J_{\pm}^iJ_{\mp}^j | \text{J}^+ \ket &= 0
&
\bra \text{J}^+ | J_{\pm}^iJ_{\mp}^j \bm{S}_i\cdot\bm{S}_j| 
\text{J}^+ \ket &= 0 \\
\bra \text{J}^+ | Q_i^2 | \text{J}^+ \ket &=
5/2
& 
\bra \text{J}^+ | \bm{S}_i\cdot\bm{S}_j| \text{J}^+ \ket &=
-3/4 \\
\notag \\
\bra \text{J}^{++} | Q_iQ_j | \text{J}^{++} \ket &= 3/4
&
\bra \text{J}^{++} | Q_iQ_j \bm{S}_i\cdot\bm{S}_j| \text{J}^{++} \ket &=-9/16 \\
\bra \text{J}^{++} | J_{\pm}^iJ_{\mp}^j | \text{J}^{++} \ket &= -3/2
&
\bra \text{J}^{++} | J_{\pm}^iJ_{\mp}^j \bm{S}_i\cdot\bm{S}_j| 
\text{J}^{++} \ket &= 9/8 \\
\bra \text{J}^{++} | Q_i^2 | \text{J}^{++} \ket &=
5/2
& 
\bra \text{J}^{++} | \bm{S}_i\cdot\bm{S}_j| \text{J}^{++} \ket &=
-3/4
\end{align}

\noindent The same observations of the $W^{1,2}$ and $W^3$ contributions as above show that these matrix elements are also consistent with gauge invariance.

The above matrix elements provide some insight into the expected hierarchy of baryon masses. In the limit $m_q\ll m_W$, the $W^\pm$ can be integrated out, and the spin-flavor eigenstates become approximate mass eigenstates. A neutral spin-flavor eigenstate can have a larger $\sum_i Q_i^2$ than a charged spin-flavor eigenstate. Then, baryon mass splittings can be dominated by quark mass splittings, and a charged baryon can be lighter than a neutral baryon. 

Even when quark mass splitting do not dominate baryon mass splittings, differences in these matrix elements can cause charged states to be lighter than neutral states when $m_q\lsim m_W$. Note for example the $Q_iQ_j$ and $J^i_\pm J^j_\mp$ matrix elements in the (2,4) model. The $W^{\pm}$ contributions to the potential cancel for the singly-charged states and not for the neutral states. So when $m_W$ is non-negligible and the electroweak inter-quark potentials are dominated by the photon, the potential is deeper for the singly-charged state than the heavier of the neutral states.

In the limit $m_q\gg m_W$, electroweak symmetry appears approximately restored in the dark sector, so the $Q_i^2$ contributions to the quark mass splittings nearly cancel, and the 5-plet baryons are approximately mass-degenerate. Then, in the effective field theory of baryons rather than quarks, electroweak self-energy corrections to baryons in a $(2J+1)$-plet are proportional to the Casimir factor $J(J+1)$. Baryons in the 5-plets therefore receive larger self-energy corrections, so they are generically expected to be heavier than the singlet baryons in the large $m_q$ limit.

As noted in \cref{subsec:IR}, neutral states in adjacent odd $\SU(2)_L$ representations (\emph{i.e.} representations whose dimensions are odd and differ by two) are forbidden from mixing because they are eigenstates of the $\mathcal{H}$-parity transformation with opposite charges. In such cases, the lack of mixing is manifest in the mass calculation because the off-diagonal spin-flavor matrix elements that would mix these states vanish exactly.

\subsection{Spatial Expectation Values}
\label{subsec:spatial}

We use the variational method to estimate the spatial expectation values in the inter-quark potential with the template wavefunction in \cref{eq:trial_psi}. We minimize baryon masses with respect to the $k$ and $k_1$ parameters, thereby placing a bound on the ground state of the Hamiltonian. We are free to use any normalizable template function with appropriate boundary conditions, and the closer the ansatz is to the true solution of the Schr\"{o}dinger equation, the better the estimate of the ground state. We chose a slight generalization of an exponentially-suppressed function, which is a typical qualitative form of bound state wavefunctions such as the ground state of the Hydrogen atom. 

Expectation values of operators $\mathcal{O}$ acting on baryon spatial states can be computed in the standard way

\begin{equation}
\langle \mathcal{O} \rangle = \frac{\int \text{d}^3r_1\cdots \text{d}^3r_{N_c} \psi^\ast \mathcal{O}\psi} {\int \text{d}^3r_1\cdots \text{d}^3r_{N_c} \psi^\ast \psi}\,.
\end{equation}

\noindent However, integrating over all positions gives a volume factor in the numerator and denominator. The integrands are translationally invariant, so it is practical to choose an origin ${\bf r}_0$ (analogous to fixing the proton at the origin in the Hydrogen atom) and enforcing it with a delta function.

\begin{equation} \label{eq:<O>}
\langle \mathcal{O} \rangle = \frac{\int \text{d}^3r_1\cdots \text{d}^3r_{N_c} \psi^\ast \mathcal{O}\psi \ \delta^{(3)}({\bf r}_0)}{\int \text{d}^3r_1\cdots \text{d}^3r_{N_c} \psi^\ast \psi \ \delta^{(3)}({\bf r}_0)}\,.
\end{equation}

\noindent The choice of origin is arbitrary, but it is convenient to choose ${\bm r}_0 = {\bm r}_{N_c}$. The required expectation values include those listed in \cref{eq:fspatial}, as well as the $\left\bra
p_i^2\right\ket$ factor in the free Hamiltonian.

The spatial integrals in \cref{eq:<O>} become increasingly challenging as $N_c$ becomes large. For the two-body wavefunction (relevant for $N_c=2$ baryons and all mesons), we analytically find 

\begin{align} \label{eq:2body_elements}
a_0 &= \frac{8(4k^2+12k_1^2+6k_1m_V+m_V^2+12kk_1+4km_V)}{(k^2+3kk_1+3k_1^2)(m_V/k+2)^5} \\
a_1 &= \frac{4k(4k^2+6k_1^2+4k_1m_V+m_V^2+8kk_1+4km_V)}{(k^2+3kk_1+3k_1^2)(m_V/k+2)^4} \\
a_2 &= \frac{4k^2(4k^2+2k_1^2+2k_1m_V+m_V^2+4kk_1+4km_V)}{(k^2+3kk_1+3k_1^2)(m_V/k+2)^3} \\
b &= \frac{4k(12k^4+20k^3m_V-6kk_1m_V^2+2km_V^3-2k_1^2m_V^2-2k_1m_V^3+2k^2k_1^2-4k^2k_1m_V+11k^2m_V^2)}{(k^2+3kk_1+3k_1^2)(m_V/k+2)^4} \\
c_1 &= -\frac{16(4k^3+8k^2k_1+4k^2m_V-3k_1^2m_V-k_1m_V^2+6kk_1^2+2kk_1m_V+km_V^2)}{(k^2+3kk_1+3k_1^2)(m_V/k+2)^5} \\
c_2 &= -\frac{8k(4k^3+4k^2k_1+4k^2m_V-2k_1^2m_V-k_1m_V^2+2kk_1^2+km_V^2)}{(k^2+3kk_1+3k_1^2)(m_V/k+2)^4} \\
d_2 &= -\frac{8k(4k^3-2kk_1m_V+km_V^2+4k^2k_1+4k^2m_V-6k_1^2m_V-2k_1m_V^2)}{(k^2+3kk_1+3k_1^2)(m_V/k+2)^5} \\
d_3 &= -\frac{4k^2(4k^3+4k^2m_V-4k_1^2m_V-2k_1m_V^2-2kk_1^2-4kk_1m_V+km_V^2)}{(k^2+3kk_1+3k_1^2)(m_V/k+2)^4} \\
\delta &= \frac{k^5}{\pi(k^2+3kk_1+3k_1^2)} \\
\left\bra p_i^2 \right\ket &= \frac{k^2(k^2+kk_1+k_1^2)}{k^2+3kk_1+3k_1^2}\,. 
\end{align}

\noindent Note that the denominators above contain identical normalization factors, and integrals whose integrands are quadratic in $\psi$ must be quadratic in $k_1$. For greater than two bodies, we resort to numerical computation. For the expectation values that do not depend on $m_V$, the dimensionful scale $k$ can be factored out of the integrand, and one can perform a scale-free numerical integral for the terms constant, linear, and quadratic in $k_1$. 

For the expectation values that do depend on $m_V$, factoring out $k$ from the integrand leaves a dependence on the ratio $m_V/k$. With some insight from the integrals we can perform analytically, we can make an ansatz for how the integrals depend on $m_V/k$. Each matrix element we have computed analytically (some of which are feasible with three bodies) organizes itself into a sum of Laurent series in $m_V/k$ with particular pole structures. The poles are at $m_V/k=-2$ for two bodies and $m_V/k=-4$ for three bodies. While unphysical, this analytic structure at negative values of $m_V/k$ helps us predict forms of the integrals for baryons with larger $N_c$. To understand why these poles appear, consider the integrand for computing $a_0$ with $k_1=0$ for three bodies

\begin{equation}
a_0 = \left\bra e^{-m_V|\bm{r}_1-\bm{r}_3|} \right\ket
\propto \int \ \text{d}^3r_1 \text{d}^3r_2 \text{d}^3r_3 \ e^{-m_V|\bm{r}_1-\bm{r}_3|} |\psi|^2\delta^{(3)}(\bm{r}_3)
\propto \int \text{d}^3u_1 \text{d}^3u_2 \ e^{-(m_V/k)u_1-2(|\bm{u}_1-\bm{u}_2|+u_1+u_2)}\,,
\end{equation}

\noindent where $\bm{u}_i=k\bm{r}_i$ is a dimensionless spatial variable. This integral is singular when the argument of the exponent has non-negative real part. If one evaluates the $\bm{u}_1$ integral while holding $\bm{u}_2$ fixed, the $m_V$-independent part of the exponent scales like $-4u_1$, and there is a singularity in the limit $m_V/k\to-4$. This argument provides some intuition as to why the poles in the analytical solutions appear at specific values of $m_V/k$. Moreover, we can extend this intuition to $N_c$ bodies, for which the $m_V$-independent part of the exponent scales like $-2(N_c-1)u_1$ holding the other variables constant, so we expect a pole in the above expectation values at $m_V/k=-2(N_c-1)$.

For $N_c>3$, the integrals may acquire additional poles at negative values of $m_V/k$ that are further from zero than $-2(N_c-1)$. This is because there are more complicated integration trajectories that give rise to a singularity. For $N_c=4$, consider 

\begin{equation}
a_0 
\propto \int \text{d}^3u_1 \text{d}^3u_2 \text{d}^3u_3 \ e^{-(m_V/k)u_1-2(|\bm{u}_1-\bm{u}_2|+|\bm{u}_1-\bm{u}_3|+|\bm{u}_2-\bm{u}_3|+u_1+u_2+u_3)}
\end{equation}

\noindent when $k_1=0$. Holding $\bm{u}_3$ fixed and integrating along the trajectory with $\bm{u}_2=\bm{u}_1$, the $m_V$-independent part of the exponent scales like $-8u_1$, which suggests a pole at $m_V/k=-8$ in addition to the pole at $m_V/k=-6$ identified above. By contrast, for the $N_c=3$ case, integrating along the trajectory $\bm{u}_2=\bm{u}_1$ suggests a pole at $m_V/k=-4$ (identical to the pole found above), so there is no reason to expect more than one pole in the three-body integrals. Likewise, integrating along the $\bm{u}_3=\bm{u}_2=\bm{u}_1$ trajectory in the four-body integral suggests the same pole at $m_V/k=-6$ as identified above. When $N_c>4$, even more poles may appear, but these become more negative and therefore have more suppressed contributions for physically relevant values of $m_V/k$ as $N_c$ increases. 

In our analysis with $N_c=4$, we estimate the $m_V$-dependence of the spatial expectation values by numerically integrating with many values of $m_V$ and performing fits to sums of Laurent series with strictly negative powers centered at each of the two expected poles

\begin{equation} \label{eq:Laurent}
\langle \mathcal{O} \rangle = \sum_{n=n_{\text{min}}}^{n_{\text{max}}} \sum_{x\in\text{poles}} \frac{f_{n,x}}{(m_V/k-x)^n}\,,
\end{equation}

\noindent where $f_{n,x}$ are fit coefficients. We do this for terms constant, linear, and quadratic in $k_1$ independently. To decide which terms to include in the Laurent series (set by $n_{\text{min}}$ and $n_{\text{max}}$ in \cref{eq:Laurent}), we again take inspiration from the two- and three-body analytical integrals. In particular, we conjecture that in the large-$m_V$ limit, where the series in \cref{eq:Laurent} is dominated by terms with $n=n_{\text{min}}$, any particular spatial matrix element decreases in magnitude with respect to $m_V$ equally quickly for all values of $N_c$. For example, $a_0\sim 1/m_V^3$ for large $m_V$ when $N_c\in\{2,3\}$, so we assume this is true for all $N_c$. This conjecture fixes $n_{\text{min}}$ and is sensible because if the decrease in $m_V$ were slower, the contribution of the term to the potential in \cref{eq:M} may \emph{increase} with $m_V$, which is clearly unphysical. To fix $n_{\text{max}}$, we observe that when $N_c$ increases by one, there is an additional radial integral in the expectation value computation, which contributes an additional factor of $r_i^2$. These two powers of $r_i$ lead to two additional powers of $m_V$ in the denominator, so we conjecture 

\begin{equation}
n_{\text{max}} = 2(N_c-2) + n_{\text{max}}^{N_c=2}\,,
\end{equation}

\noindent where $n_{\text{max}}^{N_c=2}$ is the two-body value of $n_{\text{max}}$. We find good numerical agreement with this approach.

We can check for self-consistency in our estimates of these integrals by noting that some of the spatial expectation values are inter-related by derivatives:

\begin{align}
a_0 &= -\frac{\partial\,a_1}{\partial\,m_V}
&  
a_1 &= -\frac{\partial\,a_2}{\partial\,m_V} \\
c_1 &= -\frac{\partial\,c_2}{\partial\,m_V}
& 
d_2 &= -\frac{\partial\,d_3}{\partial\,m_V}\,.
\end{align}

\noindent We found that our estimates of $a_n$ and $c_n$ agreed well with these relations. For $d_3$ with $N_c=4$, we encountered difficulties in the convergence of the numerical integrals, so we found our estimate of $d_3$ by anti-differentiating our estimate of $d_2$. Note that these derivatives are consistent with our conjectures for $n_{\text{min}}$ and $n_{\text{max}}$.

We can now express the hadron mass operator as a matrix acting on the space of hadron $\SU(2)_L$ eigenstates that depends on $k$ and $k_1$. For the variational method, we estimate the ground state by minimizing the mass with respect to $k$ and $k_1$. One remaining question is the renormalization scale $\mu$ at which to evaluate the running couplings that appear in the inter-quark potential. To answer this question, we identify the characteristic hadron length scale (the ``Bohr radius") as $1/k$, as this sets the exponential decay rate of the spatial wavefunction in \cref{eq:trial_psi}. We therefore identify the hadron's characteristic energy scale as $k$ and evaluate the running couplings at $\mu=k$. We use the one-loop renormalization group evolution of $\alpha_\chi$ in the effective field theory with the dark quarks integrated out \cite{Prosperi:2006hx}

\begin{equation} \label{eq:running_alpha}
\alpha_\chi(\mu) = \frac{6\pi}{11N_c\ln(\mu/\Lambda_\chi)}\,.
\end{equation}

\noindent We also run $\alpha_W$ and $\alpha_{\text{EM}}$ at one loop in the effective field theory without the Higgs or top quark when $\mu<m_Z$ and with the full SM when $\mu>m_Z$ (neglecting threshold effects) \cite{Alves:2014cda}. The running of $\alpha_W$ and $\alpha_{\text{EM}}$ has a very small impact. 

If one tries to minimize the mass with respect to $k$ while using $\mu=k$ in the running couplings, \cref{eq:running_alpha} shows that one encounters unphysical large logarithms that disrupt the minimization procedure. We therefore implement an iterative minimization approach. We start with an initial hypothesis for the inverse Bohr radius of $k\simeq \alpha_\chi(k)\,m_q$, which is solved by 

\begin{equation}
k\simeq \frac{6\pi \, m_q}{11N_c\,\mathcal{W}\!\left( \frac{6\pi \, m_q}{11N_c \Lambda_\chi} \right)},
\end{equation}

\noindent where $\mathcal{W}$ is the principal branch of the Lambert $W$ function (also known as the product logarithm) \cite{LambertW}. We evaluate the couplings in the inter-quark potential with $\mu$ equal to this hypothesized $k$, minimize with respect to the $k$ and $k_1$ that appear in the spatial expectation values, use the value of $k$ corresponding to that minimum as the new hypothesis for $\mu$, then repeat the minimization in this way until the minimized mass converges. This procedure ensures the couplings are evaluated at the inverse Bohr radius while avoiding large logarithms in their renormalization group evolution.

Performing the minimization procedure described in above is straightforward for hadron states that are not mixed by the mass operator (\emph{e.g.}~the charged states in the $(2,4)$ and $(4,3)$ models). These are simultaneous $\SU(2)_L$ and mass eigenstates, and each will have its own optimal $k$ and $k_1$. However, there is an ambiguity in the minimization calculation when the mass operator has off-diagonal elements. Consider the $2\times2$ mass matrix in the basis of the neutral $\SU(2)_L$ eigenstates in the $(2,4)$ or $(4,3)$ model. Each matrix element is a function of $k$ and $k_1$, as is each mass eigenvalue and the mixing angle between the $\SU(2)_L$ eigenstates. If we simply minimized each mass eigenvalue with respect to its own $k$ and $k_1$ individually, each value of the optimization parameters corresponds to a distinct set of eigenvectors. If the mass eigenstates are evaluated with different values of $k$ and $k_1$, those states are not orthogonal. In other words, the mixing angle is ill-defined if there are different values of $k$ and $k_1$ used for the two mass eigenvalues. 

We resolve this ambiguity by performing the minimization in multiple ways. We start by minimizing the lesser mass eigenvalue and using the optimal $k$ and $k_1$ values to find the greater mass eigenvalue and mass eigenstates. We then do the reverse, minimizing the greater eigenvalue and using the corresponding optimization parameters to evaluate the lesser eigenvalue and eigenstates. The differences between the masses and mixing angles computed using each method represents uncertainties in the calculation. This uncertainty in the DM mass is around $0.1\%$ for the $(2,4)$ model and $0.001\%$ for the $(4,3)$ model. The resulting uncertainty in $|\bra m_{\text{DM}}| \text{5-plet}\ket|^2$ is $\mathcal{O}(10\%)$, which shifts the steep drop-off shown in \cref{fig:mdm} by less than half an order-of-magnitude in $m_q$.

\clearpage

\bibliographystyle{utphys_modified}
\bibliography{bib}

\end{document}